\documentclass[12pt,preprint]{aastex}

\slugcomment{Accepted to appear in {\it The Astronomical Journal}}

\def\eg{{\it e.g.}}
\def\etal{{\it et al.}}
\def\etc{{\it etc.}}
\def\ie{{\it i.e.}}

\def\chisq{\chi_r^2}
\long\def\Ignore#1{\relax}

\begin{document}

\title{Photometric Decomposition of Barred Galaxies}

\author{Reese, A. S.\altaffilmark{1}, Williams, T. B.\altaffilmark{1}, Sellwood, J. A.\altaffilmark{1}, Barnes, Eric I.\altaffilmark{2}, Powell, Brian A.\altaffilmark{3}}

\altaffiltext{1}{Department of Physics and Astronomy, Rutgers University, Piscataway NJ 08854;
asreese, williams, sellwood@physics,rutgers.edu}
\altaffiltext{2}{Department of Physics, University of Wisconsin --- La Crosse, La Crosse, WI 54601}
\altaffiltext{3}{Department of Physics, University at Buffalo, The State University of New York, Buffalo, NY 14260-1500; bapowell@buffalo.edu}

\begin{abstract}
We present a non-parametric method for decomposition of the light of
disk galaxies into disk, bulge and bar components.  We have developed
and tested the method on a sample of 68 disk galaxies for which we
have acquired I-band photometry.  The separation of disk and bar light
relies on the single assumption that the bar is a straight feature
with a different ellipticity and position angle from that of the
projected disk.  We here present the basic method, but recognise that
it can be significantly refined.  We identify bars in only 47\% of the
more nearly face-on galaxies in our sample.  The fraction of light in
the bar has a broad range from 1.3\% to 40\% of the total galaxy
light.  If low-luminosity galaxies have more dominant halos, and if
halos contribute to bar stability, the luminosity functions of barred
and unbarred galaxies should differ markedly; while our sample is
small, we find only a slight difference of low significance.
\end{abstract}


\section{Introduction}
Hubble (1926) divided disk galaxies into barred and unbarred families,
denoted SB and S respectively.  A similar visual classification scheme
was described by de Vaucouleurs (1959), who extended it to include an
intermediate (SAB) family to allow for weak bars, although he noted
that the distribution of bar strengths seemed to be continuous.
Sandage \& Tamman (1981) designate intermediate types S/SB.
Subjective classifications into two or three bins might be a useful
first step, but a more quantitative estimator of the prominence of the
bar would clearly be superior.

Visual classification rests on the fact that the eye readily
distinguishes at least three separate light components of a barred
galaxy: the disk, bulge, and bar.  Ideally, we would like a reliable
procedure to separate the bar light from that of the disk and bulge in
high-quality digital images of a large sample of galaxies.  Such
information would be useful to challenge models of the origin of bars
in galaxies.  More generally, the distributions of bar light
fractions, shapes, and sizes in galaxies ought to be predicted by
theories of galaxy formation and evolution.

We will assume that the separation of the light into disk, bar, and
bulge components is well founded.  One worries whether the separate
components are dynamically distinct; \eg\ it is far from clear that
every star pursues an orbit that is always confined to the one
component.  Nevertheless, it seems reasonable to hope that the amount
of light in all three components does not change by much on an orbital
time scale and that evolution is slow.

We wish to measure a number of photometric properties of bars: the
luminosity of the bar relative to that of the disk or of the entire
galaxy, the length of the bar, the major axis light profile of the
bar, the axis ratio or ellipticity of the bar, the shape of the bar,
\etc ~Bars range from bright inner ovals (not very different from
lenses) to skinny, more ``boxy'' features (Athanassoula \etal\ 1990),
and from so-called ``flat'' bars, with a major-axis light profile that
declines slowly, to ``exponential'' bars, that have a steeply
declining light profile (Elmegreen \& Elmegreen 1985).  Such a
multi-dimensional parameter space requires something more
sophisticated than a single number to quantify the properties of a
bar.

Many attempts to characterize bars in galaxies have been reported;
early work was summarized by Sellwood \& Wilkinson (1993).  Abraham
\etal\ (1999) fit ellipses to the image at many isophote levels, and
evaluate $(b/a)^2_{\rm bar}$ from deprojection of the inner ellipses
assuming the outer ellipses indicate the projection geometry of a
round, flat disk.  Buta \& Block (2001) define bar strength $Q_b$ from
the gradients of the gravitational potential in the disk plane deduced
from a photometric image, rectified to face-on, and assuming a fixed
M/L ratio and disk thickness.  Laurikainen, Salo \& Rautianen (2002)
show that $Q_b$ correlates well with the maximum ellipticity of the
bar.  Jogee \etal\ (2004) and Marinova \& Jogee (2006) also base their
method on multiple ellipse fits, but define a galaxy to be barred from
a combination of the ellipse parameters.  Laurikainen, Salo \& Buta
(2005) quantify the non-axisymmetric light by Fourier decomposition.
Radically different approaches were adopted by Seigar \& James (1998),
who quantified the bar flux after subtraction of an axisymmetric disk,
and by Prieto \etal\ (2001), Peng \etal\ (2002), and de Souza, Gadotti
\& dos Anjos (2004) who fit assumed parametric forms for the disk,
bulge and bar to the photometric image.

Abraham \& Merrifield (2000) compare their bar-strength parameter with
optical classifications of the Frei \etal\ (1996) sample of bright
galaxies, finding some relation only when it is combined with a second
parameter characterizing the concentration of the total light.  Buta
\etal\ (2005) estimate $Q_b$ for the Eskridge \etal\ (2000) sample of
galaxies and find a fairly smooth distribution of values, suggesting a
continuum of bar strengths from zero to a maximum.  Marinova \& Jogee
(2006) reanalyse the same sample by their method, finding instead a
preponderance of strong, \ie\ quite elliptical, bars.

Most of these studies aim for a single estimator, such as bar
strength, and do not attempt to characterize the full
multi-dimensional zoo of bars.  In addition, many methods suffer from
a number of intrinsic weaknesses.

First, all require an estimate of the inclination and position angle
of the projected disk; estimating projection geometry from the outer
isophotes, assuming a flat, circular disk, can be complicated by the
presence of non-axisymmetric features, such as the bar itself as well
as spiral arms, oval distortions or warps in the outer disk.
Recognizing where such features might be important and adjusting
estimates appropriately requires practice and good judgement,
processes that are intrinsically difficult to automate.

Second, since the ellipticity at intermediate radii is affected by
spiral arms and non-circular ``rings'', as well as by bars, deciding
the radius at which a bar ends and a different bisymmetric feature
begins can be fraught with difficulty.  Recent discussions of the
appropriate way to estimate bar lengths have been given by
Athanassoula \& Misiriotis (2001), Aguerri \etal\ (2003), Erwin
(2005), Gadotti \& de Souza (2006), \etc~ Buta, Block \& Knapen (2003)
describe a Fourier method to separate the bar from the spiral before
computing $Q_b$.

Third, the ellipticity in the inner disk is affected by the bulge,
which is generally a weak contributor to the light by the end of the
bar, but usually has the highest surface brightness in the center.
Furthermore, the $Q_b$ parameter requires that the bulge be subtracted
from the disk and bar before deprojection and then added back
(Laurikainen \etal\ 2004).

Here we propose a new method to decompose the light of a galaxy in a
high-quality image.  Our method offers a more automated, and
possibly superior, approach to all three of the above difficulties,
and yields estimates of the bar light fraction, luminosity profile and
ellipticity.  We present a preliminary description here, but recognize
that there is substantial scope for improvement.  For example, we
assume an elliptical light distribution for the bar, which is
generally a poor fit; however, our method could be extended to include
more general shapes.

Our technique is a generalization of the disk-bulge decomposition
method described by Barnes \& Sellwood (2003, hereafter Paper I).  We
assume that the disk is thin, flat, and round; the bar is also thin
and flat, and is concentric with the disk.  We base our decomposition
of the bar and disk upon the assumption that the bar is straight and,
in general, has a different position angle and ellipticity from that
of the projected disk.  We make no assumptions about the light
profiles either of the disk or of the bar.  Following Palunas \&
Williams (2001), Barnes \& Sellwood also used a non-parametric form
for the bulge; here we adopt instead a more conventional S\'ersic
bulge model (S\'ersic 1968), but a parametric bulge model is not a
required feature of our three-component decompositions.

We reduce the subjectivity of defining the projection geometry by
fitting the entire light distribution at once.  Since a single
elliptical isophote can be affected by local non-axisymmetric
features, most workers generally look for a radial range over which
the ellipticity and position angle do not vary much.  By requiring a
single ellipticity for the entire disk we evaluate a weighted average
automatically.  The average ellipticity and position angle in a
disk-bulge decomposition can be thrown off by prominent
non-axisymmetric features.  Barnes \& Sellwood attempted to estimate
the systematic uncertainties caused by such features, by determining
the spread in the fitted projection angles after the image is
rectified and reprojected about other major axes.  By fitting the most
prominent non-axisymmetric feature, the bar, we significantly reduce
this potential systematic error.  The influence of bright spiral arms
remains, but they should have a lesser effect both because they are
generally less strong than bars and because they, by their nature, do
not have a fixed position angle.

We avoid judging where the bar turns to a spiral and/or ring simply by
fitting a straight feature of arbitrary light profile; the bar
intensity profile should simply drop to the noise in the outer disk.
Generally we find that the fit assigns some light to the bar beyond
the end of the bar, as judged by eye, which is probably due in part at
least to other non-axisymmetric features in the disk that are not
completely orthogonal to the bar direction.  As will be described
below, we cut off the bar light profile at some distance from the
center to prevent it from rising again where it crosses an outer
spiral arm, for example.

Finally, by fitting a separate bulge component, we are able to follow
the disk and bar light profiles in towards the center, where they
are required to have the same ellipticity and position angle as at
larger radii.

The data we use here were acquired specifically for this study.  Our
objective was to obtain high-quality images of a manageable, but
representative, sample of galaxies selected without regard to
morphological appearance.  We have used this sample to develop our
technique and present the statistics of disk, bulge, and bar light
fractions.  We hope that the method can be refined and at least partly
automated in order to be able to analyse the light of much larger
galaxy samples, \eg\ SDSS (York \etal\ 2000), the Millennium Galaxy
survey (Liske \etal\ 2006), \etc

\section{Data}
\subsection{Sample}
We have selected a magnitude-limited sample of galaxies with known
redshifts in as unbiased a manner as possible.  The redshift is
important only to determine whether the properties of the fitted bar
vary with absolute magnitude.  Our starting point was the Southern Sky
Redshift Survey (da Costa \etal\ 1998), which gives redshifts of a
complete sample of galaxies selected from the {\it Hubble Space
Telescope} Guide Star Catalog (Lasker \etal\ 1990) to a magnitude
limit of $m_{GSC} \leq 15.5$.  Da Costa \etal\ estimate the
uncertainty in the quoted magnitudes, which were determined
automatically from scans of sky survey plates, to be about 0.31
magnitudes.

We targeted a complete sub-sample of 129 objects from the SSRS in the
declination range $-16 \leq \delta \leq -14$ and right ascension
$10\;{\rm hr} \leq \alpha \leq 16\;$hr.  There were no large clusters
of galaxies among the targets, but several small groups were apparent.

\subsection{Observations}
The galaxies were observed on the nights of April 6--10, 2000 using
the CTIO 0.9 m telescope and the Tek2K\#3 CCD.  The CCD was operated
unbinned, providing an image scale of $0.396^{\prime\prime}$/pixel.
Quad readout mode was used, with an approximate gain of 3 electron/ADU
and read noise of 5 electrons per readout pixel.  The standard CTIO B,
V and I filters were used.  Observing conditions were photometric
throughout the first three nights, while the later part of the fourth
night and the entire fifth night were plagued by varying amounts of
cloud.  Seeing conditions were variable throughout the run, with image
FWHM varying from $1^{\prime\prime}$ to $3^{\prime\prime}$, with
$1.5^{\prime\prime}$ typical.  Standard stars from the fields of
Landolt (1992) were observed on the first four nights.  Twilight sky
flats were taken on the first, third, and fourth nights, and bias
frames were taken at the beginning of each night.

The observing scheme was designed to provide a deep, high-quality I
band image to be used to model the galaxy's light distribution.  We
also acquired less-deep images in the B and V bands in order to
measure the colors of the modeled components.  For each galaxy we took
two B exposures, two V exposures, and three I exposures; the telescope
was jogged slightly between exposures to minimize the effects of CCD
artifacts.  Exposure times for each image were 110 seconds in B, 75
seconds in V, and 240 seconds in I; on the last night the exposure
times were doubled.

We had time to observe 95 of the identified list of 129 target
galaxies.  The selection of galaxies to observe was based purely on
their position, without reference to any intrinsic property.  Most
of the 34 omitted galaxies were in the more southerly half of the
declination range.

\subsection{Reductions}
The images were overscan and bias subtracted, trimmed, and flattened
with the twilight sky flats in the usual way, using
IRAF.\footnote{IRAF is distributed by the National Optical Astronomy
Observatories, which are operated by the Association of Universities
for Research in Astronomy, Inc., under a cooperative agreement with
the National Science foundation.}  We determined the centroids of
stars in the images which we used to align multiple exposures; integer
pixel translations were adequate, since the pixel scale was small
compared to the seeing.  We combined the aligned images with cosmic
ray rejection.  We estimated the sky brightness in the vicinity of
each galaxy using the biweight (Beers \etal\ 1990) and subtracted it
from the images.  We also examined each image and masked out
foreground stars and any remaining artifacts.

We measured the brightnesses of standard stars in a fixed digital
aperture of diameter 2.5 times the worst seeing FWHM, and subtracted
the local sky from a surrounding annulus.  We did not attempt to
measure the atmospheric extinction, but used the standard CTIO
extinction coefficients from Stone \& Baldwin (1983).  We determined
the photometric zeropoints for the first three nights of the run, with
an internal precision of 4 to 5\%.

\section{Modeling}
We attempt to divide the light of each galaxy into distinct
components: a disk, a bar and a bulge.  Our technique is an extension
of that used for fitting a 2-D photometric image with a non-parametric
disk and bulge model, which is described in the appendix of Paper I.
While Peng \etal\ (2002) and de Souza \etal\ (2004) also fit 2-D
images, they assume the light can be decomposed into a number of
components of fixed parametric form.  Kent \& Glaudell (1989)
attempted a non-parametric fit to the light distribution of NGC~936,
but modeled the bar along its major axis only.

Here we make the usual assumption that the disk is thin, flat, and
intrinsically round; the fitted ellipticity and position angle of the
disk defines the projection geometry, therefore.  The bulge is assumed
to be spheroidal with its symmetry plane coincident with that of the
disk, and we also require the disk, bulge and bar to be concentric.
As in Paper I, we depart from convention by allowing the radial
profile of the disk light to be non-parametric, but here parameterize
the bulge as a spheroidal S\'ersic model.

The new aspect of this work is that we allow for the possibility of a
bar component in the disk, which we define to be an additional feature
in the 2-D light distribution that is bi-symmetric with no spirality.
Thus the disk and bar components are each constrained to have fixed
ellipticity and position angle at all radii, but the light profiles of
each have arbitrary tabulated values.  As will be clear from our fits,
our assumption of an elliptical shape for the bar light is generally
poor; a more boxy shape would be more appropriate.  Since we suspect
that a more sophisticated characterization of the bar light (\eg\
Athanassoula \etal\ 1990) would not have a strong effect on our
conclusions, we leave this refinement for future work.

\subsection{Mathematical details}
We minimize the reduced $\chisq$, which is defined in the usual way:
\begin{equation}
\chisq = {1 \over \nu} \sum_{i=1}^N \left( {D_i - \sum_{k=1}^K w_{k,i}
I_k - I_{s,i} \over \sigma_i} \right)^2.
\label{chisq}
\end{equation}
Here $\{D_i\}$ are the sky-subtracted intensities of the $N$ pixels
used in the image, with their associated uncertainties $\sigma_i$.
The photometric model is described by a tabulated set of intensities
$\{I_k\}$, which are interpolated to the pixel locations using the
weights $w_{k,i}$, and $I_{s,i}$ is the contribution of the model
bulge to the intensity of the $i$th pixel.

We first describe the simplest case in some detail, which is to fit
only a thin, circular disk that is inclined to the line of sight.  The
model isophotes are ellipses with fixed ellipticity $\epsilon_d$ and
position angle of the major axis $\phi_d$, all centered on the
photometric center $(x_c,y_c)$.  The intensity of the $k$th elliptical
isophote with semi-major axis $a_k$ is $I_{k,d}$.

The major-axes of all the ellipses make an angle $\phi_d$ to the
$x$-axis of the image.  We define elliptical coordinates $(x_e,y_e)$
to be the position relative to axes centered on $(x_c, y_c)$, and
rotated to be aligned with the current estimate of major-axis position
angle.  Thus a pixel center at $(x_i,y_i)$ has elliptical coordinates
\begin{eqnarray*}
x_e & = (x_i - x_c) \cos\phi_d + (y_i - y_c) \sin\phi_d  \\
y_e & = (y_i - y_c) \cos\phi_d - (x_i - x_c) \sin\phi_d.
\end{eqnarray*}
The semi-major axis, $a_i$, of the ellipse passing through this point
is given by
$$
a_i^2 = x_e^2 + \left( {y_e \over 1 - \epsilon_d} \right)^2.
$$
Since we adopt linear interpolation between the tabulated isophote
values (the data do not merit anything more sophisticated), the
two non-zero weights are
\begin{eqnarray*}
w_{k,i} & = \displaystyle{ a_{k+1} - a_i \over a_{k+1} - a_k } \\
w_{k+1,i} & = \displaystyle{ a_i - a_k \over a_{k+1} - a_k },
\end{eqnarray*}
where $\{a_k\}$ are the semi-major axes of the $K_d$ ellipses at which
the model disk intensities are tabulated.  The intensity at the central
point is $I_{1,d}$ and the model intensity for any pixel outside the
the last ellipse is computed by linear extrapolation from the
outermost pair.

Our numerical task is to find the set of parameters $\{ x_c, y_c,
\epsilon_d, \phi_d, I_{k,d}\}$ that minimize $\chisq$.  For a given
set of the parameters $(x_c, y_c, \epsilon_d, \phi_d)$, the values of
$\{I_{k,d}\}$ that minimize $\chisq$ can be obtained by solving the
linear system
\begin{equation}
\sum_{k=1}^{K_d} \left( \sum_{i=1}^N {w_{j,i}w_{k,i} \over \sigma_i^2}
\right)I_{k,d} = \sum_{i=1}^N {w_{j,i} D_i \over \sigma_i^2},
\end{equation}
obtained by setting the derivative $\partial\chisq / \partial I_j$
equal to zero.  The search for the optimum choices for $(x_c, y_c,
\epsilon_d, \phi_d)$ is by standard minimization; we adopt Powell's
method (\eg\ Press \etal\ 1992).

It is easy to include a bulge light component.  We adopt a parametric
S\'ersic bulge model
\begin{equation}
I_s(r) = I_{s,0} \exp\left\{ -b_n \left[ \left( {r \over r_e}
\right)^{1/n} -1 \right] \right\},
\label{Sersic}
\end{equation}
where $n$ is the S\'ersic index, and $I_{s,0}$ is the intensity scale.
The value of $b_n$ is set such that the effective radius, $r_e$,
contains half the light, and is defined by the implicit relation
$\Gamma(2n) = 2\gamma(2n,b_n)$, with the $\Gamma$ function and the
incomplete gamma functions having their usual definitions.

We allow the bulge to be spheroidal, with the disk plane being the
plane of symmetry.  The parameter $I_{s,0}$ adds a single extra row to
the matrix solution, whereas the bulge apparent ellipticity,
$\epsilon_s$, S\'ersic index, $n$, and effective radius, $r_e$, add
three non-linear parameters to the fit.  We do not place bounds on
these parameters, but sometimes find $n$ increases without bound at
every iteration, while $I_{s,0}$ becomes very small.  We found this
behavior to be an excellent indicator of a bulgeless galaxy.

Finally, we model the bar as an elliptical light distribution, with
ellipticity $\epsilon_b$ and projected position angle $\phi_b$ that
can differ from those of the disk.  Much of the foregoing can be
applied with ease; we add an additional set of intensities
$\{I_{k,b}\}$ to describe the bar, and all we need do is to extend the
summation $\sum_{k=1}^K w_{k,i} I_k$ such that $K=K_d + K_b$ and
$\{I_k\} = \{I_{k,d}\} \cup \{I_{k,b}\}$.  We have used the same set
of semi-major axes of the bar ellipses as for the disk, but our method
could allow the disk and bar ellipses to be spaced differently.

The $I_k$ values are not guaranteed to be positive, and we regard
significantly negative values as unacceptable, on the grounds
that they would distort the relative light fractions in the different
components.  When the best fit model includes any significantly
negative $I_k$ values, we reset those $I_k$ values to zero and re-fit.
It is sometimes necessary to do this several times to ensure that the
fitted model has no negative intensity values.  (It is possible to
enforce $I_k\geq0$ in each evaluation of $\chisq$, but this results in
discontinuities in the $\chisq$ surface as ellipses are added and
removed, which confuse the minimizing algorithm.)

\begin{figure*}
\centerline{
\includegraphics[scale=0.22]{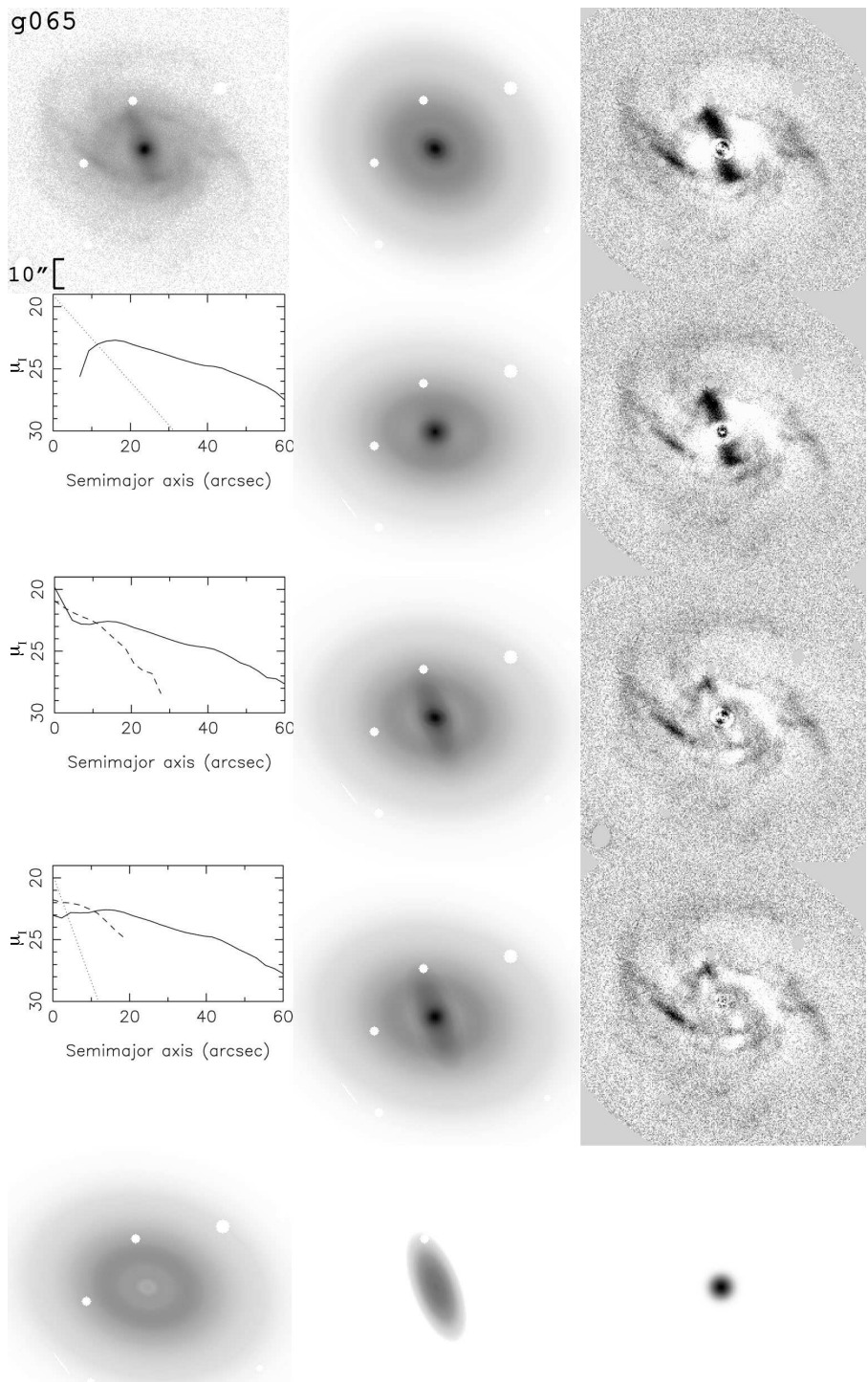}}
\caption{\footnotesize An example to illustrate our procedure in a
case with a clear bar, g065.  The top-left panel shows the I-band
image, with north up and east to the left, and the size in arcsec is
marked by the scale bar.  The other two panels in the top row show a
disk-only model fitted to the data and the residuals when this disk is
subtracted.  Each pixel in the residual maps is weighted by its
statistical uncertainty and the lighter shades indicate negative
residuals and the darker positive, the elliptical outer edge bounds
the fitted region.  Masked pixels are unshaded.  The next 3 rows show
a disk+bulge fit, a disk+bar fit and our preferred three component
disk+bulge+bar fit.  The left-hand row shows the major-axis light
profiles of the separate fitted components and the right-hand column
the residual map in each case.  Finally, the bottom row shows the
separate components from our best-fit three-component model.}
\label{steps}
\end{figure*}

\subsection{Procedure}
We selected a elliptical region of the image to be used in the
fit.  We first estimated the size and ellipticity of the galaxy by eye
and then increased the semi-major axis by 30\% and the axis ratio
$b/a$ by 10\% to ensure that all pixels containing significant galaxy
light were included.  We masked out pixels containing light from
foreground stars and other galaxies in order to eliminate them from
the fit.

We begin by fitting a disk only, and determine the four non-linear
parameters $x_c, y_c, \epsilon_d$, \& $\phi_d$.  Having found the
best-fit center, we generally hold it fixed in all subsequent fits
with additional components.  Allowing the center to float in
subsequent fits slows the fitting procedure by adding two extra
parameters, but does not lead to a significantly improved $\chisq$.
We then fit a bar and disk model, and a disk and bulge model, and
finally a three-component model with a disk, bar and bulge.

These steps are illustrated in Figure~\ref{steps} for a sample case of
the strongly barred galaxy g065 (PGC 035480).  This Figure shows the
fitted models, light profiles and residual maps from each converged
fit with the different set of components, together with the separate
components of our final fit.

The elliptical ring-like features in the centers of the residual maps
(Fig.~\ref{steps}, right hand column) for the two fits without a bulge
component are artifacts of our linear interpolation scheme.  Linear
interpolation between our somewhat coarsely spaced ellipses does not
allow the model to follow the steep bulge light profile all that well.
However, this failing is of no consequence, since the steep light
component in every galaxy is fitted by a S\'ersic bulge, leaving a
shallower light profile to be fitted by the disk and bar, for which
our interpolation scheme is entirely adequate.

Ideally, we would like an objective test to determine whether the
extra parameters of the additional components cause a significant
reduction in the value of $\chisq$, but we have not found anything
satisfactory.  We tried using the standard F-test, and also the
Bayesian information criterion (Schwarz 1978; Mukherjee \etal\ 1998),
but neither proved useful.  The number of independent pixels in our
images, after allowing for seeing, is perhaps one tenth of the actual
number, ${\cal O}(10^5$), yet is still large compared with the number
of parameters, ${\cal O}(50$), with the result that these formal tests
nearly always indicate that the extra parameters are needed, even for
a very small improvement to $\chisq$, such as might result from
fitting part of a spiral pattern with the bar model.  A further
problem is that we do not attempt to model spiral arms, dust lanes,
lop-sidedness, \etc, and therefore we find typically $1.2 \la \chisq
\la 1.8$, implying that all models are formally rejected with a high
degree of significance, which undermines the rationale of these formal
tests.  We therefore rely on visual inspection of the residuals, after
subtraction of the various models from the data, to determine the
number of separate components in the galaxy.

We have always employed equal spacing for the major axes at which the
model intensities are tabulated, even though it is not a restriction
of the procedure.  Experience revealed that thirty ellipses to
represent the disk was adequate; adding more did not reduce $\chisq$
significantly for test galaxies, but $\chisq$ was significantly worse
when many fewer were employed.  We used the same semi-major axes for
the bar, making a maximum of sixty ellipses when both components are
fitted.  The intensity profiles along a short bar are not determined
very well by only a few ellipses, and with hindsight we might have
spaced the bar ellipses more closely.  However, the principal
conclusions of this study are not affected by our undersampling of the
bar light profile.

\subsection{Discussion}
While the objective in our three component fits is to separate the
disk from the bulge and bar, this did not always happen.  In
particular, the ``bar'' light fraction sometimes turned out to be an
elliptical, presumably tri-axial, bulge misaligned with the disk major
axis.  In other cases with strong spiral arms and a weak or small bar,
the fitting software sometimes identified these outer non-bar features
as the principal non-axisymmetric component.  These latter rare cases
can be recognized by visual inspection of the model, and a new fit
undertaken with fewer bar ellipses covering the inner disk only.

The light profile of a ``bar'' that extends to the edge of the galaxy
may not decline monotonically, but rise again well outside the visual
bar where its major-axis crosses a spiral arm, for example.  In order
to find a more realistic bar, we repeat the fit with a smaller number
of bar ellipses restricted to the inner region; we retain only those
inside the first ellipse to fall below $1\sigma$ of the noise in the
first fit.  The fits that result from this procedure seem reasonable
in most cases, although the bar may be slightly larger than would have
been estimated by eye.  An alternative strategy would be to add a
penalty to the $\chisq$ function when the bar profile does not
decrease monotically; we did not employ this strategy, however, since
it would preclude the solution for the $\{I_k\}$ by linear algebra.

As our method relies on the shapes to separate the different
components, there can be significant degeneracy between them if the
shapes do not differ by much.  For example, it becomes more difficult
to separate bulge light from disk light when the galaxy is nearly
face-on as both components are then assumed to have almost circular
isophotes.  In this circumstance, our method subtracts the best-fit
S\'ersic bulge and assigns any remaining axisymmetric light to the disk.
Such degeneracies become less problematic for more inclined disks.

Here, we present our findings using the algorithm described above.  We
recognize that the light fractions in each component, and other
parameters, may be systematically biased by our procedures.  However,
it is likely that every method to quantify bars suffers from its own
idiosyncratic biases, and a method with a different set of biases is
therefore useful.  We are applying our technique to the OSUBSGS sample
(Eskridge \etal\ 2002) that has been analysed by other methods, and
will present comparisons elsewhere.

\begin{figure}[t]
\includegraphics[scale=0.21]{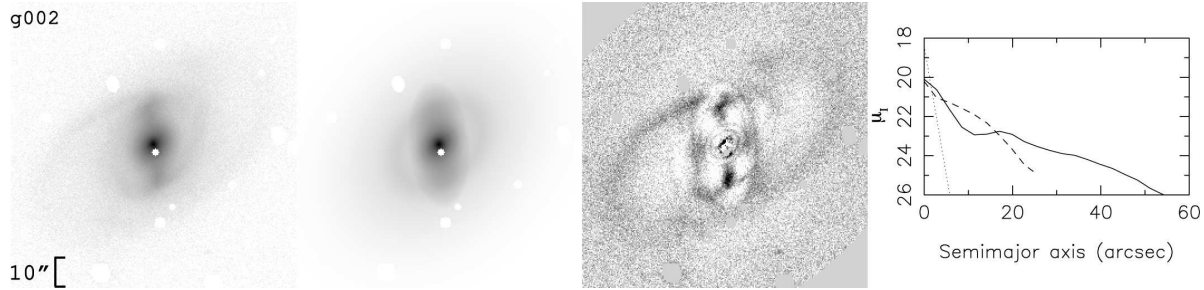}
\caption{\footnotesize The first disk galaxy in our sampe to
illustrate our fits.  The full set of 68 disk fits is given in the
on-line version.  For each galaxy, the panels from left to right show:
first, the cleaned I-band image on the left (North up and East to the
left); second, the best fit model; third, the residuals in each pixel,
weighted by its statistical uncertainty; and fourth, the surface
brightness profile of each component in the fitted model.  The scale
is in I magnitudes per square arcsec for images taken in photometric
conditions, and a simple log intensity scale otherwise.}
\label{alldisks}
\end{figure}

\section{Results}
\subsection{Magnitudes}
The apparent magnitudes of the galaxies observed during the first
three photometric nights are listed in Table 1.  The quoted values are
isophotal magnitudes to the surface brightness I=24~mag~arcsec$^{-2}$.
Comparison with existing measurements in the literature is possible in
a few cases.  Only two galaxies in our sample (g019 \& g059) have
isophotal I-band magnitudes given in the literature, and both agree
with ours to within 0.05 mag.  Since our B- \& V-band exposures were
deliberately rather short, our colors are not very precise.  We
found photographic magnitudes in the literature for 7 galaxies in the
V-band and 9 galaxies in the B-band; our estimates agree with these
published values to within their, rather large, errors.

\subsection{Decompositions}
Of the 95 galaxies observed, we exclude six spiral galaxies (22, 28,
29, 57, 61 \& 81) that had excessive light contamination from
foreground stars, 2 galaxies with images that overlap another galaxy
(64 \& 67), and two galaxies with long tails indicating mergers or
merger remnants (6 \& 62).  Throwing out these 10 cases reduces our
sample to 85 galaxies.

Our sample naturally contained elliptical galaxies.  We identified and
set aside the following 17 galaxies as ellipticals: g001, g010, g023,
g027, g031, g038, g039, g044, g045, g060, g066, g069, g078, g089,
g092, g093 \& g095.

The results of the fitting procedure are shown in
Figure~\ref{alldisks}.  From left to right, the I-band image, the best
fitted model, residual map and light profiles of the fitted components
are shown for all 68 disk galaxies.  We give the fraction of total
light in each component for the fit with the most components judged
necessary to fit the image in Table 2.  These light fractions report
the total fluxes assigned to each component in our decompositions
ignoring any flux in the masked pixels.  Including masked pixels or
extrapolating the bulge light to infinity would have hardly any effect
on the light fractions.

Uncertainties in the fluxes of the separate components were calculated
by the Metropolis-Hastings algorithm (Monte Carlo Markov chain).  We
stopped the Metropolis evaluations once the changes in the $1-\sigma$
error bars dropped below 0.03\% for an additional 20,000 accepted
points.  We found that our parameters generally have small statistical
uncertainties because of the large number of pixels.

The first disk galaxy in our sample, g002 (NGC 3469), is a good example
of a three-component model; the fitted disk, bulge (6.6\%) and bar
(28\%), illustrated in the center panel, match the original image
well.  The residuals are dominated by the prominent spiral arms in the
outer parts, but the regular pattern of residuals in the bar region
indicates that the real bar is boxier than our fitted elliptical
model.

We note galaxy g043 (NGC 3479) has a smaller bar light fraction (6\%)
and is again fitted well, with the exception of the spirals.  Galaxy
g026 (NGC 4094) has the smallest identifiable bar light fraction
(1.3\%) of any in our sample, yet the bar is clearly present.  A fit
without it has a much higher $\chisq$ and the bar stands out in the
residuals.  Galaxy g052 (NGC 4924) is a good example of an unbarred
galaxy that is fitted well by a disk and bulge only.

\begin{figure}[t]
\centerline{\includegraphics[scale=.6,angle=270]{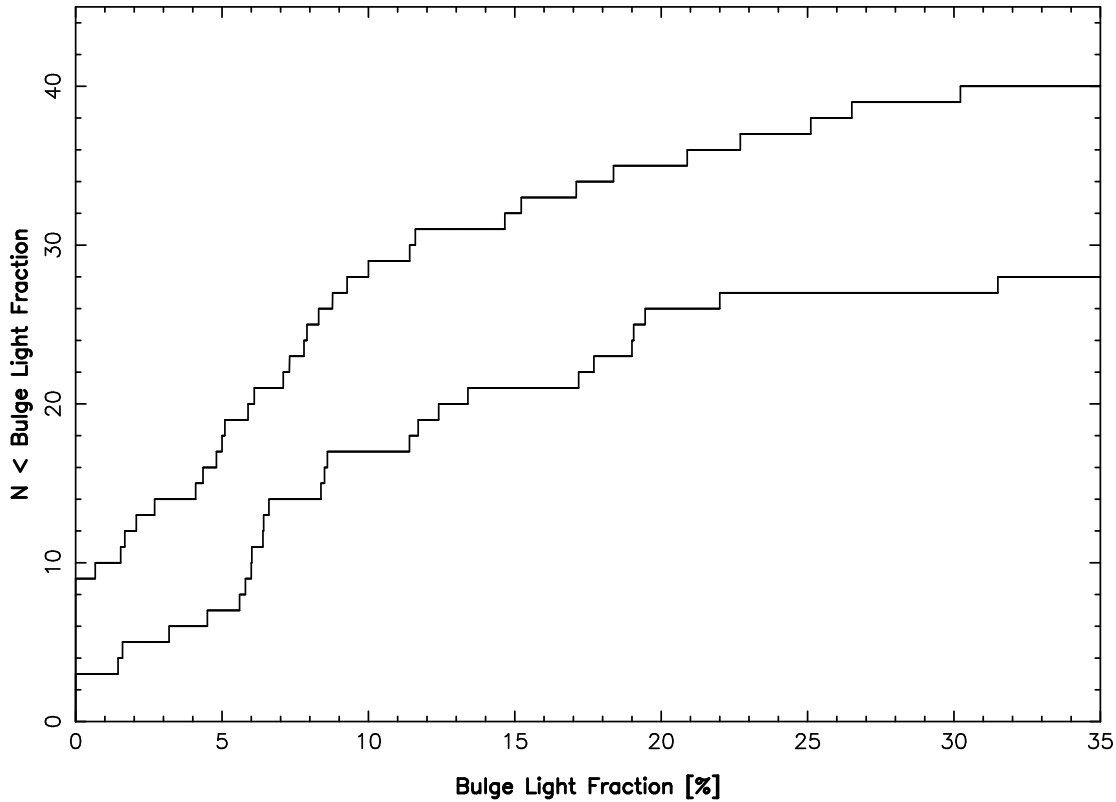}}
\caption{\footnotesize The cumulative distributions of bulge light
fractions in our sample of disk galaxies.  The lower line shows
the distribution in galaxies having identified bars, the upper line
shows that in galaxies that appear to lack bars.}
\label{bulgefracs}
\end{figure}

\subsubsection{Bulges}
Figure~\ref{bulgefracs} shows the distributions of bulge light
fractions in the barred and unbarred disks.  Of the 68 disk systems,
12 have no bulge.  The fraction of light in the bulge is generally
moderate, and is greater than 20\% in only 7 of the 68 cases, with the
largest being 31.5\%.  Apart from a slightly larger fraction of
bulgeless disks in the unbarred sample, there does not appear to be a
significant difference between the bulge light fractions in barred and
unbarred galaxies in our sample.

We find only a loose correlation, in the expected sense (\eg\ Simien
\& de Vaucouleurs 1986), between our estimated bulge light fractions
and the Hubble T parameter. \Ignore{The correlation would be tighter
if five outliers were discounted. Name them?}

\subsubsection{Bar Fraction}
As is well known, identification of bars in disks becomes more
difficult in more highly inclined galaxies.  Among the 51 galaxies
with apparent axis ratio $b/a>0.4$, we identify bars in 24, \ie\
almost half the cases.  Although we have successfully fitted bars in a
few more inclined disks, we do not find as large a bar fraction (4/17)
in these cases -- almost certainly due to this bias.  Our 47\% bar
fraction in the less inclined galaxies is on the low side of recent
estimates; the visual classification of the OSUBSGS sample by Eskridge
\etal\ (2000) finds the bar fraction to be 70\%, whereas the more
quantitative criteria employed by Laurikainen, Salo \& Buta (2004) and
Marinova \& Jogee (2006) yield bar fractions of 60\% and 62\%
respectively in the same images.

Our low bar fraction is all the more surprising since several of our
barred galaxies have $<5\%$ of the light in the bar; visual
classifications may easily miss such small features.  Furthermore, the
bars in two other cases seem to be more triaxial bulges and may
therefore have been classified visually as unbarred.  Visual
inspection of the images lacking bars by our criteria turned up just
two possible cases that might have been visually classified as barred;
in both cases, the possible bar is aligned with the inner end of a
prominent spiral pattern in nearly face-on systems.  However, it is
quite out of the question that we could have missed as many as the 11
bars that would be required to bring our bar fraction up to 70\% of
the less inclined subsample of 51.

It is possible that our sample of galaxies just happens to be
deficient in bars; the discrepancy from Eskridge \etal\ is just
$1.5\sqrt{N}$, for our $N=51$.  Apart from small numbers, the
discrepancy in bar fractions could be due to two possible systematic
differences.  First, Eskridge \etal\ used H-band images where the
identifiable fraction of bars is higher than in visual bands; it is
possible, but unlikely, that our I-band images also suffer from a
similar, but lesser, bias.  Second, the OSUBSGS is biased to higher
surface brightness, since their sample selection criteria were $B\leq
12$ and $D\leq 6^\prime$.  Perhaps bars are less common in systems of
lower surface brightness.

\begin{figure}[t]
\centerline{\includegraphics[scale=.8]{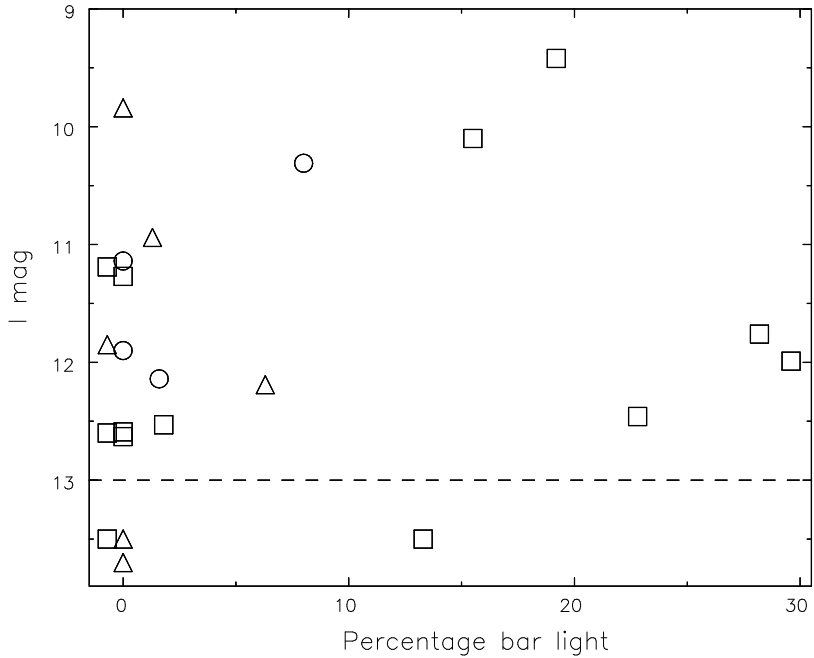}}
\caption{\footnotesize The fraction of bar light for 23 galaxies
having RC3 types.  The open circles are for galaxies classified SA,
the triangles SAB, and the squares SB.  Galaxies with no significant
bar light are shown on the left with two separate values of the
abscissae merely for clarity.  The ordinate is the I-band magnitude
where known, but is meaningless for the 4 galaxies below the
horizontal dashed line, whose images were acquired in non-photometric
conditions.}
\label{bar-unbar}
\end{figure}

\subsubsection{Comparison with Visual Classifications}
Figure~\ref{bar-unbar} shows the Hubble classifications from the RC3
for 23 disk galaxies in our sample.  We find six galaxies with SB
classifications that have no bars, while we find bars with moderate
light fractions for two galaxies that are classified SA.  The original
visual classifications were based on sky-survey images for the most
part, which are of much lower quality than our data.

\begin{figure}[t]
\centerline{\includegraphics[scale=.5,angle=270]{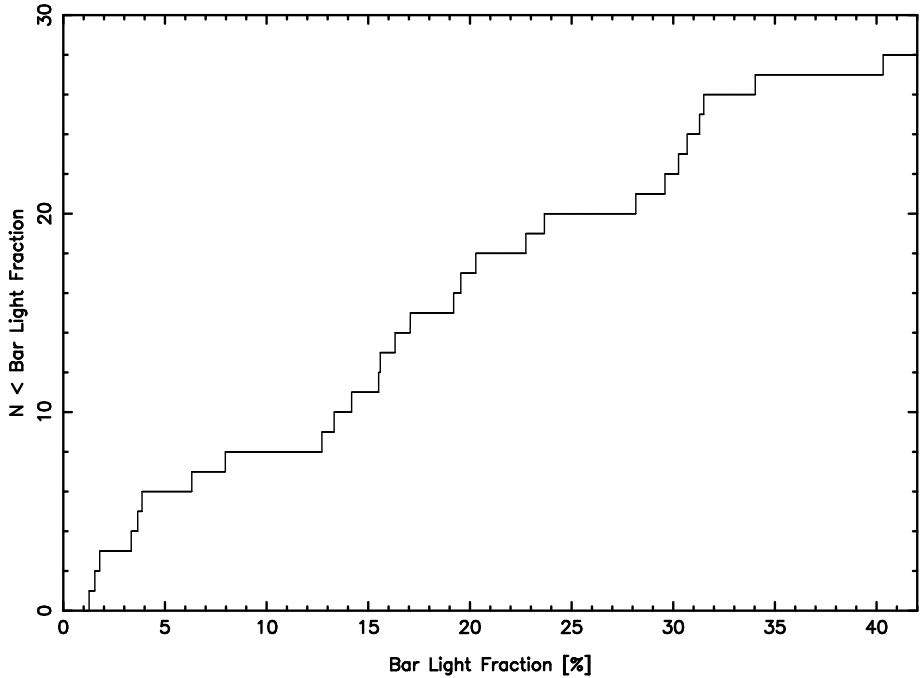}}
\caption{\footnotesize The cumulative distribution of bar light
fraction in our sample of disk galaxies.}
\label{barfracs}
\end{figure}

\subsection{Bar properties}
In all, 28 galaxies in our sample have an identifiable bar, for which
the fit was significantly improved by adding the bar component.  As
shown in Figure~\ref{barfracs}, the fraction of bar light has a broad
distribution from 1.3\% to 40\% of the total light, with no preferred
value.

We estimate the bar length to be the semi-major axis at 10\% of the
maximum of the fitted bar intensity profile.  Assuming the fitted disk
is intrinsically round, we deproject the fitted bar length and
ellipticity to find these quantities for a face-on disk.

\begin{figure}[t]
\centerline{\includegraphics[scale=.5,angle=270]{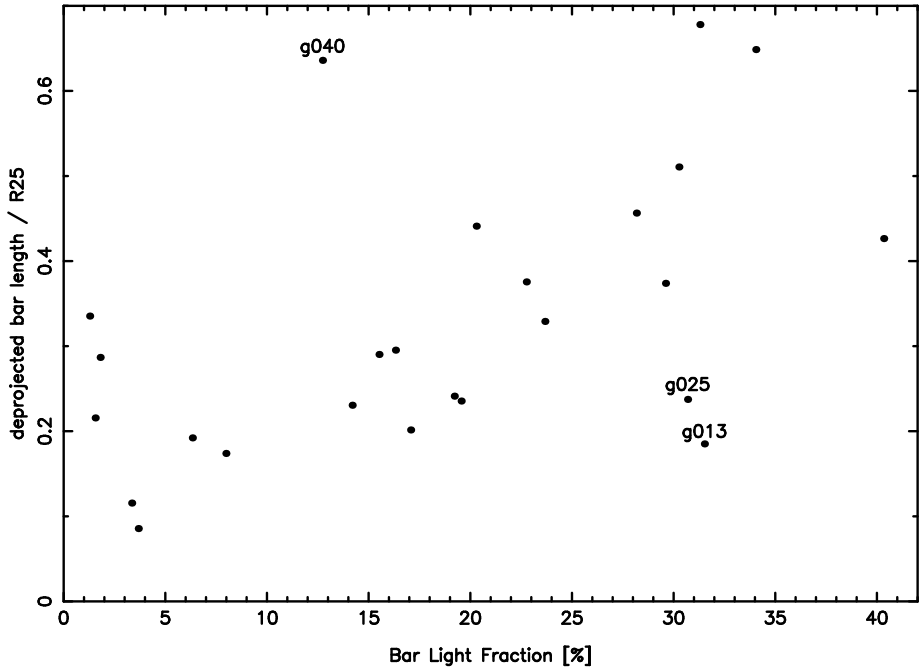}}
\caption{\footnotesize The deprojected bar length plotted as a
function of the bar light fraction in our sample of disk galaxies.}
\label{barlength}
\end{figure}

The bar light fraction correlates strongly with deprojected bar
length, expressed as a fraction of $R_{25,I}$, as shown in
Figure~\ref{barlength}.  Of the two points that lie below the trend,
g013 has a high inclination for which the deprojected the bar length
is uncertain and g025 the ``bar'' may be a triaxial bulge.  The
single outlier above the distribution, g040, has a long bar in low
surface brightness disk.  As this correlation implies a somewhat
constant surface brightness contrast for the bar, one worries whether
the absence of large, low luminosity bars is simply a selection
effect.  After re-examination of the images, we think it unlikely that
low contrast bars could have been missed, since the bars in all the
cases on the upper edge of the distribution in Fig.~\ref{barlength}
are still high-contrast features.

\begin{figure}[t]
\centerline{\includegraphics[scale=.5,angle=270]{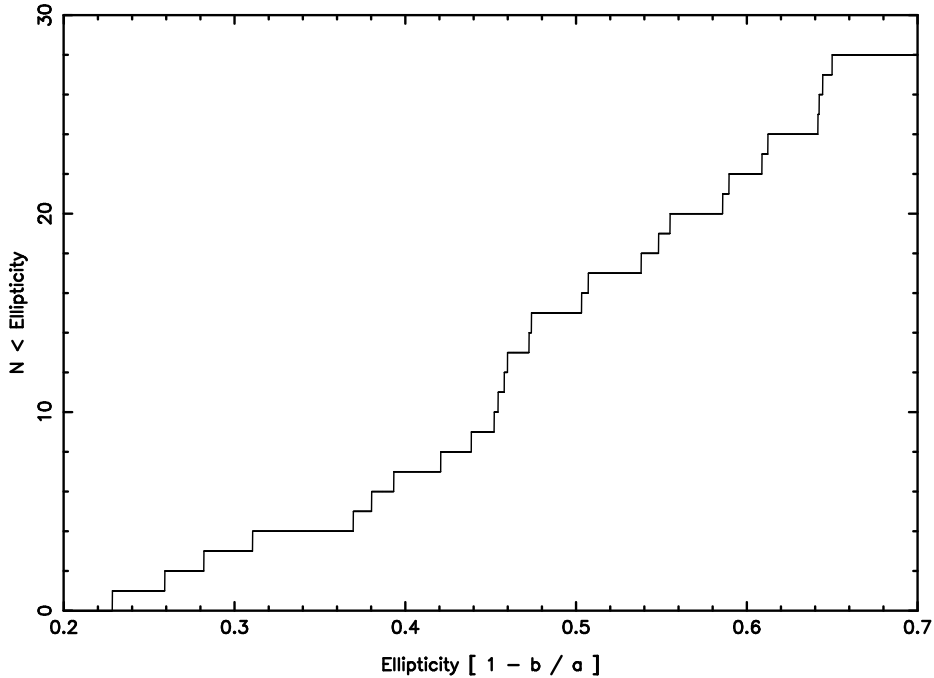}}
\caption{\footnotesize The distribution of deprojected bar
ellipticities in our sample of disk galaxies.}
\label{barevals}
\end{figure}

Figure~\ref{barevals} shows the distribution of deprojected axis
ratios, which has a broad range from $0.35 \leq b/a \leq 0.8$ (note
that we plot $1-b/a$).  Values of $b/a$ closer to unity are excluded
because such round bars would simply seem to be part of the disk.
However, the lower limit is interesting -- no bars in our sample are
skinnier than about 3:1.  Marinova \& Jogee (2006) find a similar
range of ellipticities from their different approach.  It should be
noted that this bound may depend on our use of simple ellipses; it
seems unlikely that the distribution of values shown in this Fig.\
would be affected much by use of a more general bar shape, but the
lower bound on $b/a$ may be different.

\begin{figure}[t]
\centerline{\includegraphics[scale=.9,angle=270]{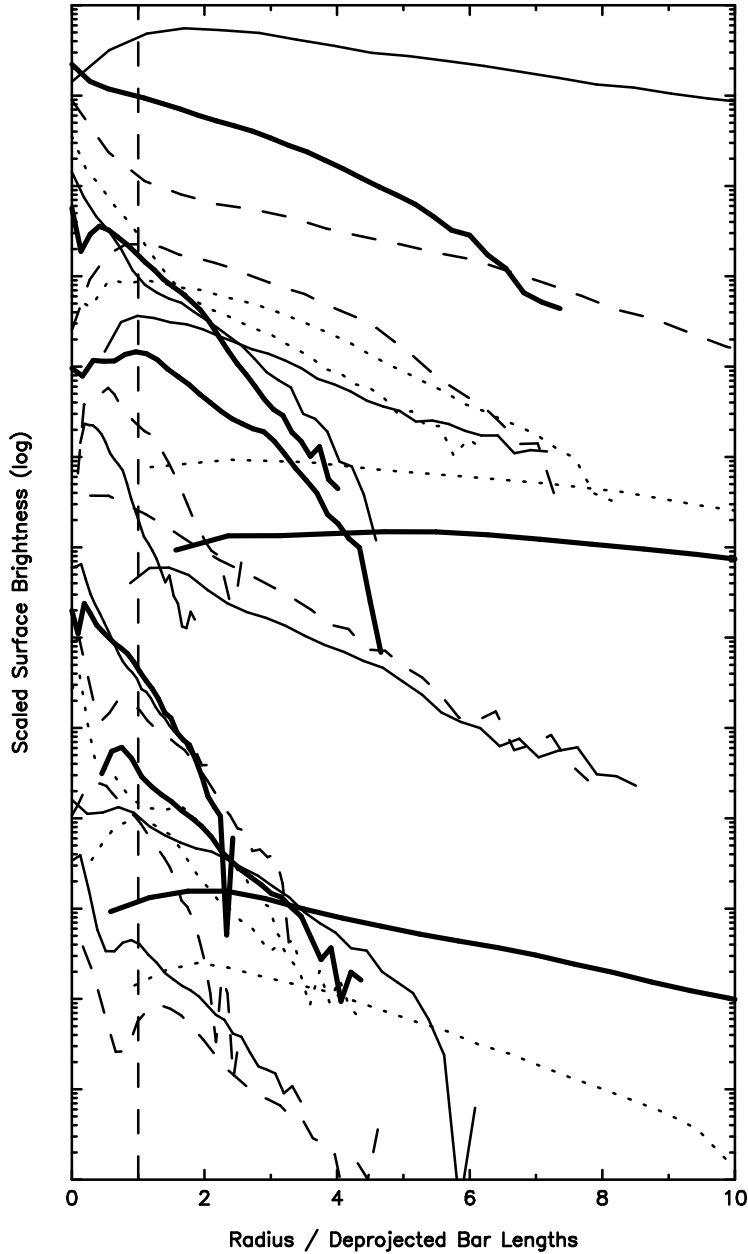}}
\caption{\footnotesize The radial light profiles of the disks of all
barred galaxies in our sample, scaled by the deprojected bar length.
The different lines are arranged from top to bottom in order of
increasing bar light fraction; they are shifted vertically by
arbitrary amounts and shown with different line styles so that they
can be easily traced.  The vertical dashed line is drawn to show the
bar length.  The disks hosting bars that contain the larger light
fractions (lower curves) generally show dips in their light profiles
in the bar region.}
\label{diskprofs}
\end{figure}

Figure~\ref{diskprofs} shows the disk light profiles, with the bar and
bulge subtracted, of all the barred galaxies in our sample.  The
galaxies are ordered from top to bottom by increasing bar light
frations.  The light profiles of many disks decrease toward the
center, which is mostly due to bulge subtraction.  However, it is
interesting that the bar component in a number of cases produces an
additional inflexion in the inner disk profile, suggesting that the
stars that make up the bar were taken from the radial range of the
bar, and are not an additional feature ``on top'' of the disk.  This
feature is generally more easily recognizable in cases with large
bar-light fractions towards the bottom of the figure.

\begin{figure}[t]
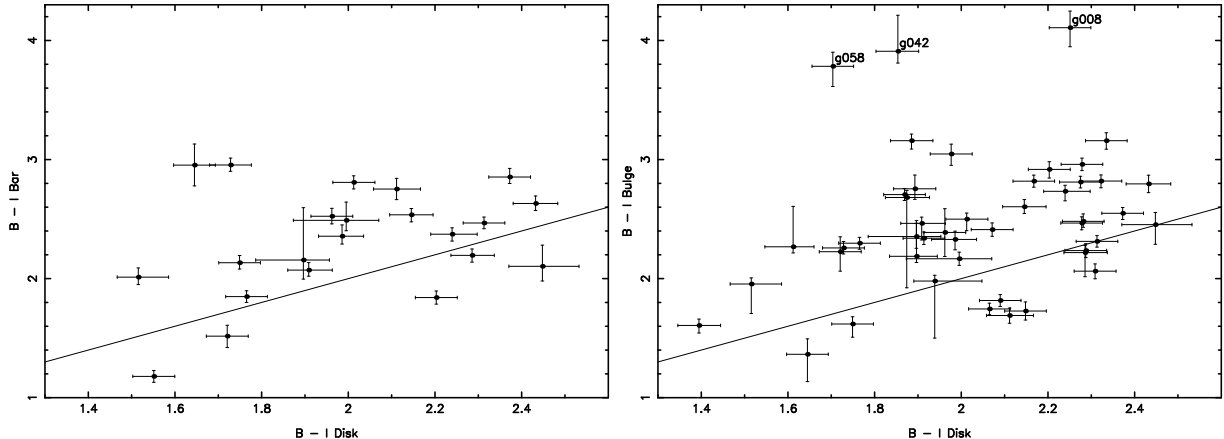

\includegraphics[scale=.33,angle=270]{bar.comp.ps}
\includegraphics[scale=.33,angle=270]{bulge.comp.ps}
\caption{\footnotesize Estimated colors of bars (above) and bulges
(below) from our model fits.  The line in each panel indicates equal
color in each component.}
\label{colcomps}
\end{figure}

\subsection{Colors}
We can estimate color differences between components by comparing
component magnitudes fitted to images taken in different filters.  As
our deeper I-band data is of higher quality, we adopt the best fit
parameters for each component from the fit to the I-band image, which
we then apply to the B and V images.  We insist that the disk, bar and
bulge have the same geometric parameters, and the same light profile
in each of the different components.  We therefore fit for the overall
normalization of the light in each component in the B- and V-bands,
using the projection parameters and light profile determined from the
best fit to the I-band image.

As shown in Figure~\ref{colcomps} (upper), we generally find bars are
slightly redder than the disk of the same galaxy and the bulges are
significantly redder still, as shown in the lower panel.  The small
color differences between the disk and bar may indicate slightly
different stellar populations, or may simply be due to rather more
dust in the bars.  We have not corrected our plotted colors for
internal extinction.

Three galaxies (g008, g042, and g058) apparently have extremely red
bulges.  All three have rather faint bulges with $<5\%$ of the total
I-band light.  One (g008) is quite edge-on, and all three appear to
have considerable internal extinction that presumably contributes to
the red colors.

\section{Noteworthy galaxies}
Two galaxies (g018 \& g021) have prominent bars in otherwise low
surface brightness disks.  The fitted bar in g018 is 34\% of the total
light, although we feel this could be an overestimate.

Our sample includes two apparently double barred galaxies: g019 (NGC
5878) and g070.  The weaker bar feature in both cases, which is easily
visible in the residual maps, appears close to perpendicular to the
bright bar and seems to be of roughly equal length!  The similar
lengths of the bright and faint bars sets them aside from the ``bars
within bars'' that are widely discussed (\eg\ Erwin \& Sparke 2002).

In addition to g026, noted above, g004 and g098 are galaxies with
extremely small bars.  Our fitting procedure could detect these very
small bars that may easily be overlooked in visual classification
schemes.

Four galaxies, g025, g030, g068 \& g096, seem to have large intrinsic
ellipticity differences between the inner and outer disks.  No one
deprojection would yield a disk with near-circular isophotes; in all
four cases the inner HSB disk appears to have a separate ellipticity
from that of the outer LSB disk.  The HSB inner disk in these cases
does not correspond to a classical lens, since is has spiral arms in
all four cases that are strongest in g030.  The existence of spirals
might suggest that the inner feature is the more nearly circular,
which would imply a highly elliptic outer disk, or one with a very
different inclination.

Our algorithm selected a compromise ellipticity for the disk light of
g025, which also has a small bar.  Curiously, the position angle of
the outer disk is within 20$^\circ$ of that of the bar.

g030 was the most difficult galaxy in our sample to fit.  Not only did
the almost round inner HSB disk have strong spirals, but the more
elongated outer disk was somewhat lop-sided.  The fitting algorithm
favored including the HSB inner disk as part of the slightly
elliptical ``bulge'' and a more elliptical disk.  There is some
evidence of a bar in the visual image, but we were unable to obtain a
3-component fit that captured this feature -- other decompositions
were preferred.

g074 and g094 are two moderately inclined disk galaxies ($i \sim
45^\circ$) that both appear to have an excess of light on the major
axis in two ``blobs'' symmetrically placed about the center.  They do
not appear as a classic bar, and may be a ring that is limb-brightened
due to projection.

\section{Discussion}
We define disk size, $R_{I,25}$, to be the semi-major axis of 25th
magnitude isophote in the I-band.  Our definition of the bar length,
$a_B$ is the semi-major axis of the bar isophote at an intensity of
10\% of its maximum.  We find little correlation between $R_{I,25}$
and $a_B$ in our models.  This disagress with Erwin's (2005) finding
that bar length scales with disk size, perhaps because we can find
much smaller bars that may have been overlooked in previous work.

We find no correlation with bar length and absolute magnitude, but
find a good correlation with disk length and absolute magnitude as
expected for systems of similar surface brightness.

Ostriker and Peebles (1973) argued that the existence of a bar in a
galaxy is determined by the dynamical importance of the dark halo;
\ie\ barred galaxies are predicted to have less dark halo than have
unbarred galaxies.\footnote{Athanassoula \& Misiriotis (2002) reported
an apparently contradictory result that bars can be excited by massive
halos.  However, such bars formed in their simulations have far too
low a pattern speed to be consistent with bars in real galaxies (\eg\
Aguerri \etal\ 2001).}  Since it is well-established that dark matter
varies inversely with galaxy luminosity, the Ostriker-Peebles
stabilization mechanism indirectly predicts a decreasing bar fraction
towards lower galaxy luminosity.  We adopt a distance to every galaxy
based on the SSRS redshift, using a Hubble constant of
$70\;\hbox{km}\;\hbox{s}^{-1}\;\hbox{Mpc}^{-1}$.

Since our sample is too small to look for a detailed trend of bar
fraction with luminosity, we simply ranked our sample of nearly
face-on galaxies by absolute luminosity.  (Eight of the 51 barred
galaxies in our sample were observed under non-photometric conditions
and therefore lack I magnitudes.)  The bar fraction in the fainter
half (8/21) is slightly lower than that in the brighter half (13/22).
However, a Kolmogorov-Smirnov test confirms that the difference in the
distributions of absolute magnitudes between the barred and unbarred
samples is of very low signficance.  If our galaxy sample is
representative, it would take a ten times larger sample for the minor
difference between the luminosity functions of barred and unbarred
galaxies to become significant.  It is already interesting that our
result is of low significance, since if bar-formation is inhibited by
massive dark halos and if halo mass fraction varies strongly with
luminosity, the luminosity functions of barred and unbarred galaxies
should differ decisively.

\section{Conclusions}
We have described a method for separating the light of a disk galaxy
into up to three separate components: a bulge, a disk, and a bar.  The
method also yields the radial light profiles of the disk and bar, as
well as the axis ratio and length of the bar.  The criterion for
separating bar light from disk light is that the two ellipitical light
components have different ellipticities and position angles that are
radially invariant.  We have not found an objective test to determine
the number of components required by the data.

We have applied this method to a sample of 68 spiral galaxies and shown
that it works well in many cases, in that the separate components
correspond to those that the eye identifies.  In these cases, the
light fractions and other properties of the separate components have
moderate statistical uncertainties.

The method is not completely objective, since it frequently fits
features in galaxies that are not at all those intended.  A triaxial
bulge can be fitted as the bar, as can bright outer spiral arms.
Lop-sided or oval disks also generally cause problems.  In most cases,
the fit can be ``nursed'' to something close to the desired
decomposition by restricting the radial range allowed for the bar fit,
although in one case (g030) we were unable to obtain a satisfacory
3-component fit despite repeated attempts with this strategy.

Our procedure is still under development.  One refinement that could
be implemented quite readily is to allow for boxiness in the bar
isophotes; the pattern of residuals after subtracting our model
frequently indicates that a better fit would be obtainable by adding
an extra bar shape parameter.  Our method also has a tendency to taper
the bar light out to larger radii than where the eye would judge the
bar to end.  It is fortunate that this weakness does not appear to
compromise our estimates of bar length, since the low surface
brightness of this excess bar light is below the 10\% threshold that
defines the bar light, but it may cause some overestimate of the bar
light fraction.

Although our statistical uncertainties are small, it is very hard to
say much in general about systematic errors.  These depend largely on
how well our model fits the image -- the fitted model matches the data
well in some cases, but the light fractions
in others are quite uncertain because the light distribution is more
complicated than our simple 3-component model.  Spiral arms are
common, and several galaxies appear not to have intrinsically round
disks; such features can, and clearly do, affect our estimates of the
light fractions in the different components, but in ways that are
impossible to quantify.

This weakness aside, we find the fraction of barred galaxies in our
sample is 47\%, which is lower than other recent estimates.  It is
unlikely that visual classifications would overestimate the bar
fraction, since our more objective scheme finds very small bars that
could easily be overlooked by eye.  Since galaxies were selected in an
unbaised manner, its lower bar fraction could simply be due to our
small sample size.

The fraction of the total light that is in a bar ranges up to some
40\%, with no preferred value.  We have identified very small bars in
a number of galaxies: the bar light in six cases is $<4\%$ of the
total, with the smallest identified fraction being 1.3\%.  The bar
light fraction correlates with the relative size of the bar,
suggesting a somewhat constant intensity contrast.

The deprojected bar axis ratios range from $0.3 \la b/a \la 0.8$.
Intrinsically rounder bars would be hard to separate from the disk,
but it is interesting that no bar was skinnier than 3:1.

We find some evidence that bars were formed from the disk, in that the
more luminous bars weakly associated with disk light profiles that dip
just interior to the bar end.  We also find that bars appear to be
redder than the host disk, but cannot say whether the color difference
is intrinsic or is due to extra dust in the bar.

We find a slightly lower fraction of bars in the fainter half of our
sample than in the brighter, but the difference of 5 cases out of 43
is barely significant.  If disks are stabilized by halos, as
originally proposed by Ostriker \& Peebles (1973), then we should
expect a lower bar fraction in low-L disks, which are known to be halo
dominated.  It is already interesting that the luminosity functions of
barred and unbarred galaxies are so similar, but a larger sample of
galaxies would provide a more decisive test of this hypothesis.

\acknowledgements We thank Roberto Abraham for providing some IDL
scripts and both him and Mike Merrifield for helpful comments on a
draft.  We also thank an anonymous referee for an extensive and useful
report.  This work was supported by NSF grants AST-0098282 \&
AST-0507323


\clearpage
\begin{deluxetable}{ccccccccc}
\tablecolumns{9}
\tablewidth{0pc}




\tablecaption{Galaxy Photometry}

\tablenum{1}

\tablehead{\colhead{Gal} & \colhead{GSC} & \colhead{Other Name} & \colhead{B \altaffilmark{b}} & \colhead{V \altaffilmark{v}} & \colhead{I \altaffilmark{i}} & \colhead{B-I} & \colhead{V-I} & \colhead{RC3 type} }

\startdata
g001 & 5505 00732 & PGC 032054              & 15.85 & 14.28 & 12.66 & 2.44 & 1.31 \\
g002 & 5506 01648 & NGC 3469                & 14.62 & 13.29 & 11.76 & 2.26 & 1.26 & PSBR2 \\
g003 & 5514 00880 & 2MASX J11183223-1426139 & 15.62 & 14.72 & 13.47 & 1.96 & 1.14 \\
g004 & 5514 00914 & 2MASX J11232225-1428148 & 15.67 & 14.48 & 13.22 & 2.17 & 1.16 \\
g005 & 5515 00365 & PGC 035176              & 15.76 & 14.42 & 12.82 & 2.32 & 1.24 \\
g006 & 5523 01245 & PGC 036744              & 16.07 & 15.34 & 14.21 & 1.42 & 0.86 & SBR6 \\
g008 &            & IC 3799                 & 14.86 & 13.84 & 12.48 & 2.14 & 1.25 & S.7*/ \\
g009 & 5541 00914 & IC 3822                 & 15.66 & 14.59 & 13.14 & 2.28 & 1.33 \\
g010 & 5541 00698 & IC 3824                 & 15.66 & 14.12 & 12.50 & 2.63 & 1.38 \\
g011 & 5541 00714 & IC 3838                 & 14.69 & 14.08 & 12.90 & 1.48 & 0.95 \\
g012 & 5541 00974 & NGC 4740                & 13.44 & 12.47 & 11.10 & 2.19 & 1.23 \\
g013 & 5542 00471 & PGC 044854              & 15.22 & 14.02 & 12.70 & 2.26 & 1.22 \\
g014 & 5550 00795 & 2MASX J13201390-1417445 & 15.43 & 14.49 & 13.21 & 2.02 & 1.17 \\
g015 & 5550 01450 & PGC 046958              & 15.06 & 14.26 & 13.07 & 1.79 & 1.09 \\
g016 & 5560 00839 & PGC 050211              & 15.25 & 14.05 & 12.59 & 2.42 & 1.35 & SBS3?/ \\
g017 & 5577 00923 & PGC 052853              & 14.93 & 13.88 & 12.60 & 1.60 & 0.98 & SBS6 \\
g018 & 5591 00907 & PGC 053654              & 15.83 & 14.57 & 12.40 & 2.25 & 1.28 \\
g019 & 5593 00656 & NGC 5878                & 12.94 & 11.72 & 10.31 & 2.34 & 1.32 & SAS3 \\
g020 & 5593 00550 & PGC 054387              & 16.56 & 14.20 & 12.47 & 2.45 & 1.35 \\
g021 & 5497 01342 & 2MASX J10255521-1443151 & 16.24 & 14.86 & 13.36 & 2.35 & 1.25 \\
g022 & 5506 00659 & 2MASX J10552639-1432581 & 15.46 & 14.87 & 13.83 & 1.44 & 0.91 \\
g023 & 5507 01145 & PGC 033775              & 15.44 & 14.06 & 12.61 & 2.34 & 1.23 \\
g024 & 5514 01149 & PGC 034225              & 15.59 & 14.24 & 12.83 & 2.33 & 1.28 & L...P \\
g025 & 5514 00991 & PGC 034459              & 15.17 & 14.19 & 12.90 & 2.04 & 1.19 \\
g026 & 5525 00637 & NGC 4094                & 13.08 & 12.33 & 10.94 & 1.69 & 1.05 & SXT6* \\
g027 & 5533 01001 & 2MASX J12295938-1437208 & 15.96 & 14.41 & 12.78 & 2.44 & 1.27 \\
g028 \altaffilmark{*} & 5533 00831  &       & 15.66 & 14.59 & 13.22 & 2.44 & 1.37 \\
g029 \altaffilmark{*} & 5534 01280 & PGC 042495 & 15.55 & 15.07 & 13.91 & 1.64 & 1.16 & IB.9?/ \\
g030 & 5534 01374 & PGC 042590              & 16.19 & 15.15 & 13.93 & 1.64 & 0.99 \\
g031 & 5541 00037 & PGC 043424              & 13.87 & 12.51 & 10.93 & 2.38 & 1.34 & RLAR+P? \\
g032 & 5541 00839 & IC 3831                 & 14.20 & 12.90 & 11.44 & 2.30 & 1.26 & PLXS0? \\
g033 & 5542 00352 & NGC 4887                & 14.26 & 13.70 & 12.74 & 1.41 & 0.90 & LA.+P? \\
g034 &            & NGC 4902                & 12.25 & 11.32 & 10.10 & 1.91 & 1.13 & SBR3 \\
g035 & 5550 01028 & IC 4221                 & 14.00 & 13.27 & 12.14 & 1.62 & 1.00 & SAR5P? \\
g036 & 5560 00352 & NGC 5420                & 14.27 & 13.30 & 12.03 & 1.98 & 1.17 & S..3* \\
g037 & 5576 00288 & NGC 5664                & 14.66 & 13.73 & 12.41 & 2.06 & 1.22 \\
g038 & 5593 00419 & NGC 5880                & 15.54 & 14.15 & 12.56 & 2.46 & 1.34 \\
g039 & 5593 00376 & NGC 5883                & 14.64 & 13.24 & 11.65 & 2.51 & 1.36 \\
g040 & 5593 00575 & 2MASX J15152969-1429317 & 15.39 & 14.41 & 13.01 & 2.04 & 1.21 \\
g041 & 5505 01314 & 2MASX J10531764-1459549 & 16.42 & 15.41 & 14.19 & 2.07 & 1.14 \\
g042 & 5506 01625 & 2MASX J10555210-1457435 & 15.46 & 14.50 & 13.46 & 1.89 & 1.14 \\
g043 & 5506 01127 & NGC 3479                & 14.78 & 13.51 & 12.19 & 1.95 & 1.14 & SXR4 \\
g044 & 5507 00560 & 2MASX J11095185-1458201 & 15.84 & 14.66 & 13.28 & 2.08 & 1.14 \\
g045 & 5507 00627 & 2MASX J11133590-1447377 & 15.75 & 14.37 & 12.87 & 2.46 & 1.30 \\
g046 & 6093 00632 & 2MASX J11521816-1500408 & 15.64 & 14.72 & 13.43 & 2.07 & 1.20 \\
g047 & 5525 00508 & PGC 038496              & 14.11 & 13.21 & 11.99 & 1.83 & 1.11 & PSBT3P* \\
g048 & 5541 00225 & PGC 044021              & 15.44 & 14.78 & 13.79 & 1.43 & 0.87 \\
g049 & 6111 00332 & PGC 044267              & 15.17 & 13.88 & 12.46 & 2.09 & 1.14 & SBT4* \\
g050 & 5542 00818 & PGC 044645              & 15.42 & 14.39 & 13.04 & 1.95 & 1.18 & L..-*/ \\
g051 & 5542 00814 & PGC 044952              & 14.48 & 13.79 & 12.62 & 1.59 & 1.00 \\
g052 & 5542 01198 & NGC 4924                & 14.10 & 13.20 & 11.85 & 1.88 & 1.13 & PSXS0P? \\
g053 & 5543 01211 & 2MASX J13153736-1452209 & 15.63 & 14.53 & 13.19 & 2.26 & 1.26 \\
g054 &            & NGC 5073                & 13.59 & 12.64 & 11.27 & 2.16 & 1.31 & SBS5?/ \\
g055 & 5550 01415 & PGC 046523              & 15.47 & 14.35 & 12.95 & 2.11 & 1.18 \\
g056 & 5551 00322 & PGC 047994              & 14.82 & 13.91 & 12.34 & 1.78 & 1.05 \\
g057 \altaffilmark{*} &  & PGC 050081       & 15.35 & 14.21 & 12.87 & 2.47 & 1.33 \\
g058 & 5560 00405 & PGC 050209              & 15.44 & 14.43 & 13.21 & 1.66 & 1.01 \\
g059 & 5577 00786 & NGC 5756                & 13.47 & 12.56 & 11.19 & 2.01 & 1.21 & PSBS4P/ \\
g060 & 5592 01120 & PGC 053889              & 15.47 & 14.14 & 12.55 & 2.53 & 1.39 \\
g061 & 6076 00771 & IRAS  10495-1504        & 15.48 & 14.86 & 13.68 & 1.57 & 0.99 \\
g062 & 6077 01161 & PGC 033374              &       &       &       &      &      \\
g063 & 6077 01600 & 2MASX J11050927-1521039 & 15.93 & 15.25 & 14.16 & 1.52 & 0.92 \\
g064 & 6077 00021 & PGC 034046              & 15.48 & 14.73 & 13.69 & 1.60 & 0.95 \\
g065 & 6085 00590 & PGC 035480              &       &       &       &      &      & PSBR3 \\
g066 & 6093 00259 & PGC 037281              &       &       &       &      &      \\
g067 \altaffilmark{*} & 6111 00448 & PGC 043547 & 15.68 & 14.15 & 12.40 & 2.47 & 1.36 \\
g068 & 6111 00192 & PGC 043625              & 14.89 & 13.93 & 12.47 & 2.07 & 1.23 \\
g069 & 6111 00630 & NGC 4756                & 13.85 & 12.49 & 11.04 & 2.35 & 1.29 & LXS0? \\
g070 &            & NGC 4856                & 11.90 & 10.78 &  9.42 & 2.34 & 1.31 & SBS0 \\
g071 & 6112 00735 & PGC 044701              & 14.78 & 13.92 & 12.63 & 1.80 & 1.07 & IBS9?/ \\
g072 & 6112 01482 & NGC 4877                & 13.80 & 12.52 & 11.14 & 2.16 & 1.26 & SAS2* \\
g073 & 6113 00188 & PGC 045958              &       &       &       &      &     & SXS8 \\
g074 & 6113 01709 & PGC 046436              &       &       &       &      &      \\
g075 & 6122 00596 & PGC 048102              & 15.10 & 14.58 & 12.42 & 2.25 & 1.84 \\
g076 & 6122 00579 & PGC 048144              & 16.26 & 15.89 & 13.34 & 1.83 & 1.65 \\
g077 & 6123 00184 & 2MASX J13445714-1515151 &       &       &       &      &      \\
g078 & 6153 00233 & PGC 052002              &       &       &       &      &      \\
g079 & 6154 00667 & PGC 052840              &       &       &       &      &      \\
g080 & 6168 01004 & PGC 053634              &       &       &       &      &      & SBS7 \\
g081 & 6062 00839 & 2MASX J10380158-1531057 &       &       &       &      &      \\
g088 & 5525 00508 & PGC 038496              & 14.77 & 13.72 & 12.53 & 1.80 & 1.04 & PSBT3P* \\
g089 \altaffilmark{*} & & PGC 043646        & 16.63 & 15.25 & 13.75 & 2.45 & 1.34 \\
g090 & 6111 01092 & PGC 043664              & 15.10 & 14.25 & 12.84 & 1.92 & 1.24 \\
g091 & 6111 01258 & IC 0829                 & 14.62 & 13.56 & 12.35 & 2.10 & 1.12 \\
g092 & 6111 01164 & 2MASX J12523303-1531010 & 15.65 & 14.38 & 12.80 & 2.30 & 1.30 \\
g093 & 6111 00632 & PGC 043777              & 15.88 & 14.20 & 12.83 & 2.31 & 1.09 \\
g094 &            & PGC 044471              & 14.04 & 13.14 & 11.97 & 2.00 & 1.14 \\
g095 & 6112 00084 & 2MASX J13071600-1543396 & 15.80 & 15.10 & 12.84 & 2.40 & 1.83 \\
g096 &            & NGC 4984                & 12.35 & 11.26 &  9.84 & 2.22 & 1.27 & RLXT+ \\
g098 & 6113 00989 & PGC 046334              &       &       &       &      &      \\
g112 & 6095 00798 & PGC 038529              & 14.54 & 13.30 & 11.90 & 2.63 & 1.54 & SAS1? \\
g113 & 6104 01587 & 2MASX J12390630-1610472 & 15.38 & 14.37 & 13.09 & 1.99 & 1.13 \\
g120 & 6113 01387 & PGC 046350              &       &       &       &      &      \\
g124 & 6121 00362 & 2MASX J13331239-1608216 &       &       &       &      &      \\
g125 & 6121 00343 & PGC 47717               &       &       &       &      &      & SXS2P* \\

\enddata


\tablecomments{B,V,and I are R25 magnitudes. B-I and V-I are calculated at the I25 isophote.}
\tablenotetext{b}{B uncertainty is +/- 0.048 mags}
\tablenotetext{v}{V uncertainty is +/- 0.035 mags}
\tablenotetext{i}{I uncertainty is +/- 0.046 mags}
\tablenotetext{*}{galaxy heavily contaminated by nearby star}
\end{deluxetable}
\clearpage

\clearpage
\def\arraystretch{1.500}
\begin{deluxetable}{ccccccccccc}
\rotate
\tablewidth{0pc}
\tablecaption{Galaxy Decompositions}
\tablenum{2}
\tablehead{\colhead{Name} & \colhead{disk \% \altaffilmark{a}} & \colhead{bar \% \altaffilmark{a}} & \colhead{bulge \% \altaffilmark{a}} & \colhead{$\epsilon_d$} & \colhead{$\phi_d$} & \colhead{$\epsilon_b$} & \colhead{$\phi_b$} & \colhead{$\epsilon_s$} & \colhead{$n$} & \colhead{$r_e$}}
\startdata
g002 	& $	65.3	^{+	0.8	}_{-	0.3	}$ & $	28.2	^{+	0.3	}_{-	0.8	}$ & $	6.6	^{+	0.2	}_{-	0.2	}$ & $	0.296	^{+	0.001	}_{-	0.012	}$ & $	138.9	^{+	0.2	}_{-	1.4	}$ & $	0.53	^{+	0.01	}_{-	0.01	}$ & $	84.6	^{+	0.3	}_{-	0.3	}$ & $	0.01	^{+	0.01	}_{-	0.01	}$ & $	0.83	^{+	0.02	}_{-	0.03	}$ & $	3.4	^{+	0.1	}_{-	0.1	}$ \\
g003 	& $	88.5	^{+	0.1	}_{-	3.5	}$ &						 & $	11.5	^{+	3.4	}_{-	0.2	}$ & $	0.32	^{+	0.009	}_{-	0.001	}$ & $	124.6	^{+	0.2	}_{-	0.7	}$ &						&						& $	0.25	^{+	0.03	}_{-	0.06	}$ & $	0.59	^{+	0.09	}_{-	0.04	}$ & $	4.4	^{+	0.3	}_{-	0.3	}$ \\
g004 	& $	74.5	^{+	0.4	}_{-	0.3	}$ & $	17.1	^{+	0.4	}_{-	0.4	}$ & $	8.5	^{+	0.3	}_{-	0.4	}$ & $	0.518	^{+	0.004	}_{-	0.005	}$ & $	42.9	^{+	0.3	}_{-	0.3	}$ & $	0.3	^{+	0.01	}_{-	0.01	}$ & $	7.7	^{+	1.9	}_{-	2.1	}$ & $	0.07	^{+	0.03	}_{-	0.03	}$ & $	0.73	^{+	0.03	}_{-	0.05	}$ & $	2.1	^{+	0.1	}_{-	0.1	}$ \\
g005 	& $	82.9	^{+	0.5	}_{-	0.4	}$ & $	3.7	^{+	0.1	}_{-	0.1	}$ & $	13.4	^{+	0.4	}_{-	0.5	}$ & $	0.281	^{+	0.006	}_{-	0.008	}$ & $	54.6	^{+	0.8	}_{-	0.5	}$ & $	0.5	^{+	0.02	}_{-	0.02	}$ & $	110.9	^{+	1.3	}_{-	1.5	}$ & $	0.01	^{+	0.01	}_{-	0.01	}$ & $	1.03	^{+	0.05	}_{-	0.06	}$ & $	3.8	^{+	0.2	}_{-	0.3	}$ \\
g008 	& $	97.9	^{+	0.1	}_{-	0.1	}$ &						 & $	2.1	^{+	0.1	}_{-	0.1	}$ & $	0.867	^{+	0.001	}_{-	0.001	}$ & $	58.1	^{+	0.1	}_{-	0.1	}$ &						&						& $	0.46	^{+	0.03	}_{-	0.02	}$ & $	0.63	^{+	0.08	}_{-	0.02	}$ & $	6.2	^{+	0.1	}_{-	0.1	}$ \\
g009 	& $	93.9	^{+	0.1	}_{-	0.2	}$ &						 & $	6.1	^{+	0.2	}_{-	0.1	}$ & $	0.826	^{+	0.002	}_{-	0.001	}$ & $	54.7	^{+	0.1	}_{-	0.1	}$ &						&						& $	0.21	^{+	0.06	}_{-	0.03	}$ & $	0.92	^{+	0.07	}_{-	0.05	}$ & $	5.2	^{+	0.2	}_{-	0.1	}$ \\
g011 	& $	73.5	^{+	1	}_{-	0.6	}$ &						 & $	26.5	^{+	0.6	}_{-	1.1	}$ & $	0.002	^{+	0.009	}_{-	0.003	}$ & $	91.7	^{+	1.3	}_{-	0.6	}$ &						&						& $	0.23	^{+	0.02	}_{-	0.02	}$ & $	1.19	^{+	0.07	}_{-	0.08	}$ & $	12.9	^{+	0.4	}_{-	0.4	}$ \\
g012 	& $	77.7	^{+	0.7	}_{-	0.2	}$ & $	16.3	^{+	0.4	}_{-	0.4	}$ & $	6	^{+	0.1	}_{-	0.5	}$ & $	0.25	^{+	0.002	}_{-	0.003	}$ & $	148.5	^{+	0.3	}_{-	0.3	}$ & $	0.28	^{+	0.01	}_{-	0.02	}$ & $	39.7	^{+	1	}_{-	0.7	}$ & $	0.01	^{+	0.01	}_{-	0.01	}$ & $	0.85	^{+	0.01	}_{-	0.04	}$ & $	3.9	^{+	0	}_{-	0.1	}$ \\
g013 	& $	62.8	^{+	1.1	}_{-	0.3	}$ & $	31.5	^{+	0.6	}_{-	0.9	}$ & $	5.6	^{+	0.2	}_{-	1.1	}$ & $	0.731	^{+	0.004	}_{-	0.004	}$ & $	92.9	^{+	0.2	}_{-	0.2	}$ & $	0.27	^{+	0.01	}_{-	0.03	}$ & $	102.6	^{+	2.3	}_{-	1.1	}$ & $	0.21	^{+	0.02	}_{-	0.06	}$ & $	0.61	^{+	0.14	}_{-	0.02	}$ & $	2.6	^{+	0.3	}_{-	0.1	}$ \\
g014 	&	100					&						&						& $	0.173	^{+	0.005	}_{-	0.007	}$ & $	72.3	^{+	1.2	}_{-	0.9	}$ &						&						&						&						&						\\
g015 	& $	74.1	^{+	0.2	}_{-	0.2	}$ & $	19.6	^{+	0.6	}_{-	0.2	}$ & $	6.4	^{+	0.1	}_{-	0.6	}$ & $	0.323	^{+	0.005	}_{-	0.006	}$ & $	2.4	^{+	0.5	}_{-	0.4	}$ & $	0.56	^{+	0.01	}_{-	0.01	}$ & $	166.6	^{+	0.4	}_{-	0.3	}$ & $	0.02	^{+	0.03	}_{-	0.03	}$ & $	0.31	^{+	0.03	}_{-	0.04	}$ & $	5.7	^{+	0.1	}_{-	0.2	}$ \\
g016 	& $	69.8	^{+	0.1	}_{-	0.2	}$ &						 & $	30.2	^{+	0.2	}_{-	0.1	}$ & $	0.787	^{+	0.002	}_{-	0.002	}$ & $	118.7	^{+	0.1	}_{-	0.1	}$ &						&						& $	0.28	^{+	0.01	}_{-	0.01	}$ & $	1.65	^{+	0.02	}_{-	0.02	}$ & $	9.4	^{+	0.1	}_{-	0.1	}$ \\
g017 	&	100					&						&						& $	0.238	^{+	0.012	}_{-	0.006	}$ & $	149.4	^{+	0.9	}_{-	1.4	}$ &						&						&						&						&						\\
g018 	& $	43.9	^{+	2.5	}_{-	4.9	}$ & $	34	^{+	4.9	}_{-	2.4	}$ & $	22	^{+	0.6	}_{-	0.6	}$ & $	0.001	^{+	0.002	}_{-	0.001	}$ & $	136.8	^{+	0.7	}_{-	0.4	}$ & $	0.23	^{+	0.02	}_{-	0.03	}$ & $	42.4	^{+	0.9	}_{-	1	}$ & $	0.23	^{+	0.02	}_{-	0.02	}$ & $	0.75	^{+	0.01	}_{-	0.02	}$ & $	4.8	^{+	0.1	}_{-	0.1	}$ \\
g019 	& $	73	^{+	0.2	}_{-	0.3	}$ & $	8	^{+	0.4	}_{-	0.1	}$ & $	19	^{+	0.1	}_{-	0.4	}$ & $	0.649	^{+	0.001	}_{-	0.001	}$ & $	86.3	^{+	0.1	}_{-	0.1	}$ & $	0.6	^{+	0.01	}_{-	0.01	}$ & $	101.4	^{+	0.3	}_{-	0.5	}$ & $	0.31	^{+	0.01	}_{-	0.01	}$ & $	1.21	^{+	0.01	}_{-	0.06	}$ & $	9.6	^{+	0.1	}_{-	0.1	}$ \\
g020 	& $	62.9	^{+	0.9	}_{-	0.9	}$ & $	31.3	^{+	0.9	}_{-	0.8	}$ & $	5.8	^{+	0.3	}_{-	0.3	}$ & $	0.265	^{+	0.007	}_{-	0.01	}$ & $	157	^{+	1.1	}_{-	0.8	}$ & $	0.59	^{+	0.01	}_{-	0.01	}$ & $	4.6	^{+	0.3	}_{-	0.3	}$ & $	0.21	^{+	0.01	}_{-	0.02	}$ & $	0.67	^{+	0.03	}_{-	0.04	}$ & $	2.7	^{+	0.1	}_{-	0.1	}$ \\
g021 	& $	60.6	^{+	4.1	}_{-	3.5	}$ & $	20.3	^{+	3.4	}_{-	2.2	}$ & $	19.1	^{+	1.7	}_{-	3	}$ & $	0.338	^{+	0.022	}_{-	0.012	}$ & $	116.1	^{+	1	}_{-	3.8	}$ & $	0.52	^{+	0.06	}_{-	0.02	}$ & $	114.1	^{+	1.9	}_{-	0.8	}$ & $	0.22	^{+	0.05	}_{-	0.09	}$ & $	0.62	^{+	0.15	}_{-	0.06	}$ & $	3.5	^{+	0.9	}_{-	0.3	}$ \\
g024 	& $	48.3	^{+	1.5	}_{-	0.4	}$ & $	40.3	^{+	0.8	}_{-	1.2	}$ & $	11.4	^{+	0.1	}_{-	0.7	}$ & $	0.54	^{+	0.008	}_{-	0.006	}$ & $	95.8	^{+	0.3	}_{-	0.5	}$ & $	0.44	^{+	0.01	}_{-	0.01	}$ & $	70.4	^{+	0.9	}_{-	1	}$ & $	0.19	^{+	0.02	}_{-	0.02	}$ & $	0.75	^{+	0.03	}_{-	0.03	}$ & $	3.9	^{+	0.1	}_{-	0.2	}$ \\
g025 	& $	51.6	^{+	1.3	}_{-	0.1	}$ & $	30.7	^{+	1.2	}_{-	0.4	}$ & $	17.7	^{+	0.3	}_{-	2.3	}$ & $	0.53	^{+	0.008	}_{-	0.008	}$ & $	15.6	^{+	0.5	}_{-	0.4	}$ & $	0.26	^{+	0.01	}_{-	0.04	}$ & $	183.4	^{+	0.7	}_{-	2.8	}$ & $	0.01	^{+	0.01	}_{-	0.01	}$ & $	0.79	^{+	0.02	}_{-	0.04	}$ & $	4.3	^{+	0.1	}_{-	0.2	}$ \\
g026 	& $	94.2	^{+	0.6	}_{-	0.5	}$ & $	1.3	^{+	0.1	}_{-	0.1	}$ & $	4.5	^{+	0.5	}_{-	0.7	}$ & $	0.639	^{+	0.001	}_{-	0.002	}$ & $	25.9	^{+	0.1	}_{-	0.1	}$ & $	0.55	^{+	0.02	}_{-	0.02	}$ & $	48.3	^{+	2.3	}_{-	2.8	}$ & $	0.54	^{+	0.02	}_{-	0.03	}$ & $	0.51	^{+	0.03	}_{-	0.04	}$ & $	18.6	^{+	0.1	}_{-	0.1	}$ \\
g030 	& $	88.4	^{+	0.1	}_{-	3.9	}$ &						 & $	11.6	^{+	3.9	}_{-	0.1	}$ & $	0.14	^{+	0.008	}_{-	0.006	}$ & $	5.9	^{+	1.2	}_{-	2	}$ &						&						& $	0.01	^{+	0.01	}_{-	0.01	}$ & $	0.77	^{+	0.03	}_{-	0.02	}$ & $	4	^{+	0.1	}_{-	0.1	}$ \\
g032 	& $	90	^{+	0.2	}_{-	0.2	}$ &						 & $	10	^{+	0.2	}_{-	0.3	}$ & $	0.388	^{+	0.002	}_{-	0.002	}$ & $	123	^{+	0.2	}_{-	0.1	}$ &						&						& $	0.23	^{+	0.02	}_{-	0.01	}$ & $	0.88	^{+	0.02	}_{-	0.02	}$ & $	4.8	^{+	0.1	}_{-	0.1	}$ \\
g033 	& $	74.9	^{+	0.5	}_{-	0.7	}$ &						 & $	25.1	^{+	0.7	}_{-	0.5	}$ & $	0.472	^{+	0.005	}_{-	0.004	}$ & $	113.2	^{+	0.3	}_{-	0.2	}$ &						&						& $	0.14	^{+	0.04	}_{-	0.03	}$ & $	0.94	^{+	0.03	}_{-	0.02	}$ & $	8.5	^{+	0.2	}_{-	0.2	}$ \\
g034 	& $	72.1	^{+	0.1	}_{-	0.4	}$ & $	15.5	^{+	0.5	}_{-	0.3	}$ & $	12.4	^{+	0.3	}_{-	0.2	}$ & $	0.021	^{+	0.001	}_{-	0.012	}$ & $	-9.1	^{+	9.1	}_{-	9.5	}$ & $	0.65	^{+	0.01	}_{-	0.01	}$ & $	23.1	^{+	0.3	}_{-	0.1	}$ & $	0.18	^{+	0.01	}_{-	0.01	}$ & $	1.31	^{+	0.01	}_{-	0.09	}$ & $	13.7	^{+	0	}_{-	0.2	}$ \\
g035 	& $	97.6	^{+	0.3	}_{-	0.1	}$ & $	1.6	^{+	0.1	}_{-	0.1	}$ & $	0.9	^{+	0.1	}_{-	0.2	}$ & $	0.522	^{+	0.003	}_{-	0.003	}$ & $	100.6	^{+	0.2	}_{-	0.2	}$ & $	0.68	^{+	0.02	}_{-	0.03	}$ & $	129	^{+	3.5	}_{-	1.8	}$ & $	0.07	^{+	0.07	}_{-	0.07	}$ & $	0.65	^{+	0.07	}_{-	0.17	}$ & $	3.9	^{+	0	}_{-	0.4	}$ \\
g036 	& $	95.7	^{+	0.3	}_{-	0.2	}$ &	 					 & $	4.4	^{+	0.2	}_{-	0.4	}$ & $	0.604	^{+	0.001	}_{-	0.001	}$ & $	133.3	^{+	0	}_{-	0	}$ &						&						& $	0.27	^{+	0.06	}_{-	0.06	}$ & $	1.25	^{+	0.08	}_{-	0.07	}$ & $	8.1	^{+	0.5	}_{-	0.8	}$ \\
g037 	& $	81.6	^{+	0.2	}_{-	0.3	}$ &	  					 & $	18.4	^{+	0.3	}_{-	0.2	}$ & $	0.566	^{+	0.002	}_{-	0.002	}$ & $	58.3	^{+	0.2	}_{-	0.1	}$ &						&						& $	0.08	^{+	0.03	}_{-	0.03	}$ & $	1.14	^{+	0.03	}_{-	0.03	}$ & $	5.7	^{+	0.1	}_{-	0.1	}$ \\
g040 	& $	63.9	^{+	1.6	}_{-	1.3	}$ & $	12.7	^{+	0.6	}_{-	0.5	}$ & $	23.3	^{+	0.9	}_{-	1	}$ & $	0.443	^{+	0.013	}_{-	0.012	}$ & $	74.7	^{+	0.9	}_{-	0.8	}$ & $	0.75	^{+	0.01	}_{-	0.01	}$ & $	68.8	^{+	0.4	}_{-	0.4	}$ & $	0.19	^{+	0.01	}_{-	0.01	}$ & $	1.03	^{+	0.05	}_{-	0.07	}$ & $	4.3	^{+	0.2	}_{-	0.3	}$ \\
g041 	& $	77.3	^{+	0.9	}_{-	0.4	}$ &						 & $	22.7	^{+	0.4	}_{-	0.9	}$ & $	0.639	^{+	0.004	}_{-	0.012	}$ & $	74.8	^{+	0.3	}_{-	0.3	}$ &						&						& $	0.18	^{+	0.07	}_{-	0.02	}$ & $	0.92	^{+	0.04	}_{-	0.1	}$ & $	4.8	^{+	0.2	}_{-	0.3	}$ \\
g042 	& $	94.9	^{+	0.1	}_{-	1.5	}$ &						 & $	5.1	^{+	1.5	}_{-	0.1	}$ & $	0.54	^{+	0.002	}_{-	0.003	}$ & $	82.5	^{+	0.1	}_{-	0.2	}$ &						&						& $	0.12	^{+	0.1	}_{-	0.04	}$ & $	0.5	^{+	0.06	}_{-	0.04	}$ & $	3.2	^{+	0.2	}_{-	0.1	}$ \\
g043 	& $	92.1	^{+	0.4	}_{-	0.4	}$ & $	6.3	^{+	0.2	}_{-	0.2	}$ & $	1.6	^{+	0.3	}_{-	0.3	}$ & $	0.282	^{+	0.005	}_{-	0.004	}$ & $	86.5	^{+	0.4	}_{-	0.3	}$ & $	0.41	^{+	0.02	}_{-	0.02	}$ & $	129.4	^{+	1.2	}_{-	1.2	}$ & $	0.11	^{+	0.06	}_{-	0.06	}$ & $	0.58	^{+	0.06	}_{-	0.1	}$ & $	3.8	^{+	0.5	}_{-	0.6	}$ \\
g046 	& $	92.2	^{+	0.6	}_{-	0.5	}$ &	  					& $	7.8	^{+	0.5	}_{-	0.6	}$ & $	0.38	^{+	0.004	}_{-	0.003	}$ & $	196.5	^{+	0.3	}_{-	0.3	}$ &						&						& $	0.05	^{+	0.1	}_{-	0.06	}$ & $	0.89	^{+	0.07	}_{-	0.08	}$ & $	3.5	^{+	0.2	}_{-	0.3	}$ \\
g047 	& $	57.4	^{+	0.1	}_{-	0.1	}$ & $	29.6	^{+	0.3	}_{-	0.1	}$ & $	13	^{+	0.1	}_{-	0.2	}$ & $	0.383	^{+	0.006	}_{-	0.003	}$ & $	39.5	^{+	0.3	}_{-	0.2	}$ & $	0.66	^{+	0.01	}_{-	0.01	}$ & $	44.9	^{+	0.1	}_{-	0.1	}$ & $	0.04	^{+	0.01	}_{-	0.02	}$ & $	0.93	^{+	0.02	}_{-	0.02	}$ & $	5.3	^{+	0.1	}_{-	0.1	}$ \\
g048 	& $	69.7	^{+	0.3	}_{-	0.3	}$ & $	30.3	^{+	0.3	}_{-	0.3	}$ &	  					& $	0.542	^{+	0.009	}_{-	0.012	}$ & $	147.8	^{+	0.7	}_{-	0.8	}$ & $	0.62	^{+	0.01	}_{-	0.01	}$ & $	118.7	^{+	0.7	}_{-	0.8	}$ &						&						&						\\
g049 	& $	73.1	^{+	0.9	}_{-	1.8	}$ & $	22.8	^{+	1.8	}_{-	0.9	}$ & $	4.2	^{+	0.2	}_{-	0.3	}$ & $	0.32	^{+	0.005	}_{-	0.006	}$ & $	114.8	^{+	0.8	}_{-	0.5	}$ & $	0.44	^{+	0.02	}_{-	0.02	}$ & $	151.4	^{+	1.1	}_{-	1.6	}$ & $	0.1	^{+	0.03	}_{-	0.03	}$ & $	0.72	^{+	0.03	}_{-	0.05	}$ & $	2.3	^{+	0.1	}_{-	0.1	}$ \\
g050 	&	100					&	 					&						& $	0.697	^{+	0.003	}_{-	0.003	}$ & $	74.3	^{+	0.2	}_{-	0.2	}$ &						&						&						&						&						\\
g051 	&	100					&	 					&						& $	0.422	^{+	0.004	}_{-	0.005	}$ & $	101.8	^{+	0.4	}_{-	0.4	}$ &						&						&						&						&						\\
g052 	& $	95.2	^{+	0.1	}_{-	0.1	}$ &	 					 & $	4.8	^{+	0.1	}_{-	0.1	}$ & $	0.191	^{+	0.003	}_{-	0.003	}$ & $	32.7	^{+	0.5	}_{-	0.3	}$ &						&						& $	0.08	^{+	0.02	}_{-	0.01	}$ & $	0.62	^{+	0.02	}_{-	0.03	}$ & $	3.6	^{+	0.1	}_{-	0.1	}$ \\
g053 	& $	92.9	^{+	0.2	}_{-	0.2	}$ &	 					 & $	7.1	^{+	0.2	}_{-	0.2	}$ & $	0.639	^{+	0.002	}_{-	0.002	}$ & $	77.5	^{+	0.1	}_{-	0.2	}$ &						&						& $	0.25	^{+	0.06	}_{-	0.07	}$ & $	0.77	^{+	0.06	}_{-	0.03	}$ & $	5.5	^{+	0.2	}_{-	0.2	}$ \\
g054 	& $	92.7	^{+	0.1	}_{-	0.1	}$ &	 					 & $	7.3	^{+	0.1	}_{-	0.1	}$ & $	0.844	^{+	0.001	}_{-	0.001	}$ & $	120.3	^{+	0.1	}_{-	0.1	}$ &						&						& $	0.47	^{+	0.01	}_{-	0.01	}$ & $	0.88	^{+	0.02	}_{-	0.01	}$ & $	9.4	^{+	0.1	}_{-	0.1	}$ \\
g055 	& $	85.3	^{+	0.4	}_{-	0.4	}$ &	  					 & $	14.7	^{+	0.4	}_{-	0.4	}$ & $	0.497	^{+	0.007	}_{-	0.007	}$ & $	209.9	^{+	0.3	}_{-	0.3	}$ &						&						& $	0.09	^{+	0.04	}_{-	0.03	}$ & $	1.21	^{+	0.06	}_{-	0.06	}$ & $	4.2	^{+	0.2	}_{-	0.2	}$ \\
g056 	& $	64.9	^{+	3.3	}_{-	0.5	}$ & $	23.7	^{+	1.4	}_{-	0.9	}$ & $	11.5	^{+	0.2	}_{-	2.8	}$ & $	0.081	^{+	0.016	}_{-	0.006	}$ & $	-38	^{+	38	}_{-	3	}$ & $	0.42	^{+	0.02	}_{-	0.02	}$ & $	121.3	^{+	0.8	}_{-	0.3	}$ & $	0.22	^{+	0.04	}_{-	0.01	}$ & $	0.58	^{+	0.06	}_{-	0.02	}$ & $	2.7	^{+	0.1	}_{-	0.1	}$ \\
g058 	& $	97.3	^{+	0.1	}_{-	0.1	}$ &	  					 & $	2.7	^{+	0.1	}_{-	0.1	}$ & $	0.14	^{+	0.009	}_{-	0.012	}$ & $	57.9	^{+	2.1	}_{-	1.8	}$ &						&						& $	0.01	^{+	0.01	}_{-	0.01	}$ & $	0.65	^{+	0.01	}_{-	0.06	}$ & $	3.9	^{+	0.1	}_{-	0.1	}$ \\
g059 	& $	92.1	^{+	0.1	}_{-	0.1	}$ &	  					 & $	7.9	^{+	0.1	}_{-	0.1	}$ & $	0.581	^{+	0.001	}_{-	0.001	}$ & $	48.5	^{+	0.1	}_{-	0.1	}$ &						&						& $	0.13	^{+	0.01	}_{-	0.01	}$ & $	1.15	^{+	0.01	}_{-	0.02	}$ & $	4.4	^{+	0.1	}_{-	0.1	}$ \\
g063 	&	100					&	 					&						& $	0.639	^{+	0.006	}_{-	0.008	}$ & $	141	^{+	0.5	}_{-	0.5	}$ &						&						&						&						&						\\
g065 	& $	79.2	^{+	0.3	}_{-	0.4	}$ & $	13.3	^{+	0.2	}_{-	0.4	}$ & $	7.5	^{+	0.2	}_{-	0.1	}$ & $	0.241	^{+	0.007	}_{-	0.005	}$ & $	13.1	^{+	1.2	}_{-	0.2	}$ & $	0.59	^{+	0.01	}_{-	0.02	}$ & $	69.5	^{+	0.6	}_{-	0.3	}$ & $	0.01	^{+	0.01	}_{-	0.01	}$ & $	0.68	^{+	0.03	}_{-	0.01	}$ & $	4.9	^{+	0.1	}_{-	0.1	}$ \\
g068 	& $	96.7	^{+	0.1	}_{-	0.1	}$ & $	3.3	^{+	0.1	}_{-	0.1	}$ &						& $	0.266	^{+	0.004	}_{-	0.004	}$ & $	192.9	^{+	0.6	}_{-	0.5	}$ & $	0.45	^{+	0.02	}_{-	0.01	}$ & $	16.5	^{+	1.1	}_{-	1.2	}$ &						&						&						\\
g070 	& $	49.3	^{+	0.1	}_{-	0.1	}$ & $	19.2	^{+	0.2	}_{-	0.1	}$ & $	31.5	^{+	0.1	}_{-	0.1	}$ & $	0.647	^{+	0.001	}_{-	0.001	}$ & $	51.4	^{+	0.1	}_{-	0.1	}$ & $	0.48	^{+	0.01	}_{-	0.01	}$ & $	35	^{+	0.3	}_{-	0.2	}$ & $	0.3	^{+	0.01	}_{-	0.01	}$ & $	1.33	^{+	0.01	}_{-	0.01	}$ & $	17.2	^{+	0.1	}_{-	0.1	}$ \\
g071 	&	100					&	 					&						& $	0.677	^{+	0.003	}_{-	0.003	}$ & $	140.9	^{+	0.2	}_{-	0.2	}$ &						&						&						&						&						\\
g072 	& $	95	^{+	0.1	}_{-	0.1	}$ &	  					 & $	5	^{+	0.1	}_{-	0.1	}$ & $	0.572	^{+	0.001	}_{-	0.001	}$ & $	82.7	^{+	0.1	}_{-	0.1	}$ &						&						& $	0.17	^{+	0.02	}_{-	0.01	}$ & $	0.92	^{+	0.02	}_{-	0.02	}$ & $	3.9	^{+	0.1	}_{-	0.1	}$ \\
g073	& $	99.3	^{+	0.1	}_{-	0.1	}$ &						 & $	0.7	^{+	0.1	}_{-	0.1	}$ & $	0.001	^{+	0.001	}_{-	0.001	}$ & $	80.1	^{+	12	}_{-	8.7	}$ &						&						& $	0.15	^{+	0.08	}_{-	0.06	}$ & $	0.67	^{+	0.09	}_{-	0.07	}$ & $	4.7	^{+	0.4	}_{-	0.5	}$ \\
g074 	& $	90.7	^{+	0.1	}_{-	0.1	}$ &	 					 & $	9.3	^{+	0.1	}_{-	0.1	}$ & $	0.715	^{+	0.001	}_{-	0.001	}$ & $	165.7	^{+	0.1	}_{-	0.1	}$ &						&						& $	0.35	^{+	0.02	}_{-	0.01	}$ & $	0.89	^{+	0.03	}_{-	0.03	}$ & $	6.3	^{+	0.1	}_{-	0.1	}$ \\
g075 	& $	92.5	^{+	0.2	}_{-	0.2	}$ &	 					 & $	7.5	^{+	0.2	}_{-	0.2	}$ & $	0.448	^{+	0.003	}_{-	0.002	}$ & $	150.8	^{+	0.2	}_{-	0.1	}$ &						&						& $	0.04	^{+	0.05	}_{-	0.04	}$ & $	0.88	^{+	0.03	}_{-	0.03	}$ & $	4.1	^{+	0.1	}_{-	0.1	}$ \\
g076	& $	95.9	^{+	0.6	}_{-	1.5	}$ &						 & $	4.1	^{+	0.5	}_{-	1.8	}$ & $	0.151	^{+	0.012	}_{-	0.015	}$ & $	100.4	^{+	1.5	}_{-	1.9	}$ &						&						& $	0.18	^{+	0.05	}_{-	0.03	}$ & $	0.59	^{+	0.12	}_{-	0.25	}$ & $	3.9	^{+	1.3	}_{-	1.8	}$ \\
g077 	&	100					&	 					&						& $	0.145	^{+	0.015	}_{-	0.024	}$ & $	83.3	^{+	4.4	}_{-	3.5	}$ &						&						&						&						&						\\
g079 	& $	77.2	^{+	2.1	}_{-	0.4	}$ & $	15.6	^{+	0.9	}_{-	2.2	}$ & $	7.3	^{+	0.8	}_{-	0.8	}$ & $	0.327	^{+	0.009	}_{-	0.006	}$ & $	154.4	^{+	0.8	}_{-	0.6	}$ & $	0.43	^{+	0.04	}_{-	0.03	}$ & $	180.6	^{+	3.1	}_{-	0.8	}$ & $	0.17	^{+	0.03	}_{-	0.02	}$ & $	0.82	^{+	0.01	}_{-	0.14	}$ & $	3.7	^{+	0.2	}_{-	0.4	}$ \\
g080 	&	100					&	 					&						& $	0.666	^{+	0.024	}_{-	0.001	}$ & $	143.8	^{+	1	}_{-	0.6	}$ &						&						&						&						&						\\
g088 	& $	98.2	^{+	0.3	}_{-	0.2	}$ & $	1.8	^{+	0.2	}_{-	0.3	}$ &						& $	0.071	^{+	0.003	}_{-	0.01	}$ & $	39.4	^{+	2.4	}_{-	0.2	}$ & $	0.62	^{+	0.03	}_{-	0.05	}$ & $	27.1	^{+	3.2	}_{-	4.4	}$ &						&						&						\\
g090	& $	91.2	^{+	0.5	}_{-	0.9	}$ &						& $	8.8	^{+	0.9	}_{-	0.6	}$ & $	0.369	^{+	0.006	}_{-	0.004	}$ & $	109.8	^{+	0.3	}_{-	0.3	}$ &						&						& $	0.21	^{+	0.05	}_{-	0.08	}$ & $	0.74	^{+	0.03	}_{-	0.04	}$ & $	4.2	^{+	0.3	}_{-	0.4	}$ \\
g091	& $	64.9	^{+	1.9	}_{-	6.5	}$ & $	14.2	^{+	6.2	}_{-	2.1	}$ & $	20.9	^{+	2.1	}_{-	1.8	}$ & $	0.199	^{+	0.01	}_{-	0.008	}$ & $	156.4	^{+	1.7	}_{-	1.2	}$ & $	0.29	^{+	0.04	}_{-	0.08	}$ & $	116.9	^{+	9.6	}_{-	3.4	}$ & $	0.24	^{+	0.01	}_{-	0.01	}$ & $	0.61	^{+	0.04	}_{-	0.05	}$ & $	4	^{+	0.2	}_{-	0.2	}$ \\
g094 	& $	79.1	^{+	0.1	}_{-	0.1	}$ &	 					 & $	20.9	^{+	0.1	}_{-	0.1	}$ & $	0.691	^{+	0.001	}_{-	0.001	}$ & $	178.5	^{+	0.1	}_{-	0.1	}$ &						&						& $	0.3	^{+	0.01	}_{-	0.01	}$ & $	0.84	^{+	0.01	}_{-	0.01	}$ & $	4.5	^{+	0.1	}_{-	0.1	}$ \\
g096 	& $	98.3	^{+	0.1	}_{-	0.1	}$ &	 					 & $	1.7	^{+	0.1	}_{-	0.1	}$ & $	0.134	^{+	0.001	}_{-	0.001	}$ & $	23.2	^{+	0.2	}_{-	0.2	}$ &						&						& $	0.01	^{+	0.01	}_{-	0.01	}$ & $	0.39	^{+	0.01	}_{-	0.01	}$ & $	2.7	^{+	0.1	}_{-	0.1	}$ \\
g098 	& $	84.4	^{+	0.8	}_{-	0.5	}$ & $	3.9	^{+	0.2	}_{-	0.3	}$ & $	11.7	^{+	0.2	}_{-	0.8	}$ & $	0.465	^{+	0.003	}_{-	0.003	}$ & $	115.6	^{+	0.2	}_{-	0.1	}$ & $	0.47	^{+	0.02	}_{-	0.02	}$ & $	147.7	^{+	1.7	}_{-	1.5	}$ & $	0.17	^{+	0.04	}_{-	0.03	}$ & $	0.91	^{+	0.03	}_{-	0.03	}$ & $	4.5	^{+	0.1	}_{-	0.1	}$ \\
g112 	& $	84.8	^{+	0.2	}_{-	0.2	}$ &	  					 & $	15.2	^{+	0.6	}_{-	0.2	}$ & $	0.503	^{+	0.003	}_{-	0.003	}$ & $	106.3	^{+	0.1	}_{-	0.1	}$ &						&						& $	0.18	^{+	0.02	}_{-	0.02	}$ & $	0.85	^{+	0.01	}_{-	0.02	}$ & $	4.9	^{+	0.1	}_{-	0.1	}$ \\
g113 	& $	82.8	^{+	8.1	}_{-	0.2	}$ &	 					 & $	17.2	^{+	0.1	}_{-	8.2	}$ & $	0.082	^{+	0.052	}_{-	0.072	}$ & $	180.5	^{+	1.1	}_{-	3.1	}$ &						&						& $	0.03	^{+	0.33	}_{-	0.09	}$ & $	0.67	^{+	0.02	}_{-	0.27	}$ & $	4.2	^{+	15.6	}_{-	13.2	}$ \\
g120 	& $	98.5	^{+	0.2	}_{-	0.2	}$ &	 					 & $	1.5	^{+	0.1	}_{-	0.2	}$ & $	0.586	^{+	0.001	}_{-	0.001	}$ & $	62.8	^{+	0	}_{-	0	}$ &						&						& $	0.34	^{+	0.12	}_{-	0.13	}$ & $	0.58	^{+	0.07	}_{-	0.08	}$ & $	3.4	^{+	0.4	}_{-	13.2	}$ \\
g124 	&	100					&	 					&						& $	0.154	^{+	0.017	}_{-	0.02	}$ & $	150.4	^{+	3.6	}_{-	2.9	}$ &						&						&						&						&						\\
g125 	& $	94.1	^{+	0.1	}_{-	0.1	}$ &	  					 & $	5.9	^{+	0.1	}_{-	0.1	}$ & $	0.506	^{+	0.001	}_{-	0.006	}$ & $	153.1	^{+	0.1	}_{-	0.3	}$ &						&						& $	0.19	^{+	0.03	}_{-	0.02	}$ & $	0.98	^{+	0.03	}_{-	0.05	}$ & $	4.5	^{+	0.1	}_{-	0.2	}$ \\

\enddata
\tablenotetext{a}{Disk, bar, and bulge refer to the fraction of light in each component.}
\end{deluxetable}

\end{document}